\documentstyle[amssymb,preprint,aps]{revtex}
\tightenlines
\begin{document}
\title{ADIABATIC FOLLOWING CRITERION, ESTIMATION OF THE NONADIABATIC
EXCITATION FRACTION AND QUANTUM JUMPS}
\author{R. N. Shakhmuratov,$^{1,2}$ and J. Odeurs$^{1}$ }
\address{$^{1}$Instituut voor Kern- en Stralingsfysica, Katholieke\\
Universiteit Leuven, Celestijnenlaan 200 D, B-3001 Leuven, Belgium\\
$^{2}$Kazan Physical-Technical Institute, Russian Academy of Sciences, 10/7\\
Sibirsky Trakt Street, Kazan 420029 Russia}
\maketitle

\begin{abstract}
An accurate theory describing adiabatic following of the dark, nonabsorbing
state in the three-level system is developed. An analytical solution for the
wave function of the particle experiencing Raman excitation is found as an
expansion in terms of the time varying nonadiabatic perturbation parameter.
The solution can be presented as a sum of adiabatic and nonadiabatic parts.
Both are estimated quantitatively. It is shown that the limiting value to
which the amplitude of the nonadiabatic part tends is equal to the Fourier
component of the nonadiabatic perturbation parameter taken at the Rabi
frequency of the Raman excitation. The time scale of the variation of both
parts is found. While the adiabatic part of the solution varies slowly and
follows the change of the nonadiabatic perturbation parameter, the
nonadiabatic part appears almost instantly, revealing a jumpwise transition
between the dark and bright states. This jump happens when the nonadiabatic
perturbation parameter takes its maximum value.
\end{abstract}

\section{INTRODUCTION}

Stimulated Raman adiabatic passage (STIRAP) resulting in the population
transfer between the states which are not directly coupled \cite{Bergmann98}%
, electromagnetically induced transparency (EIT) via adiabatic following of
the dark, non-absorbing state \cite{Shakhmuratov2001}, nonresonant pulse
excitation of the two-level atom \cite{Crisp1973} are just a short list of
phenomena in quantum optics where adiabatic processes are considered.
Generally, one can find in any part of physics problems concerned with
adiabaticity, which are treated almost similarly. Among them we can mention
multiphoton resonances induced in atoms and molecules by a strong
low-frequency field \cite{Keldysh64}-\cite{Dykhne1970}, wave-packet dynamics
in physics and chemistry or, so-called, ''femto-chemistry'' \cite{Zewail93}-%
\cite{Suominen95} and slow atomic and molecular collisions \cite{Landau32}-%
\cite{Demkov69}.

A fully adiabatic process takes place if the Hamiltonian of the problem is
diagonal. Then any time dependence of this diagonal Hamiltonian does not
change the state of the quantum system except its phase. In general, the
Hamiltonian, describing the interacting systems, has both diagonal and
nondiagonal components. In most cases dealing with adiabatic excitation, the
problem can be reduced to the consideration of the two-level system
perturbed by the diagonal and nondiagonal interactions. Then, both
components of the Hamiltonian, i.e., the two-level splitting and the
nondiagonal perturbation, are time dependent. One can find instantaneous
eigenstates of such a Hamiltonian. These states are called adiabatic states
if the system, being initially in one of them, follows this state. In
contrast, the initial states, where the transition takes place, are called
diabatic states.

The diagonalization of the instantaneous Hamiltonian takes into account the
diabatic states coupling, incorporating the interaction parameter into the
instantaneous eigenvalues. However, a new coupling appears again because of
the time dependence of the transformation from the initial, diabatic basis
to the adiabatic basis (see, for example, \cite{Suominen95}). Inasmuch as
the coupling between instantaneous eigenstates does not vanish, the
following of the adiabatic states can never be perfect. Since it is almost
impossible to get rid of the state coupling by the time dependent
transformation of the basis functions, the choice of the adiabatic basis is
aimed to minimize the transition probability between the states, almost
approaching in this sense to the fully adiabatic Hamiltonian (i.e., the
diagonal one) when the coupling can be neglected. The estimation of the
fraction of the system that is transferred from one adiabatic state to
another during the interaction (nonadiabatic correction) is a main problem
in this approach. This nonadiabatic correction is a measure of the
nonadiabaticity of the process.

It should be obvious that the searching for the adiabatic basis is an
approximate approach. It was developed for the first time to treat the
behavior of the two-level system subject to an excitation that couples the
states if both the two-level splitting and the excitation are time
dependent. Only a few models of such a two-level system have analytical
solutions. Among them are the Rosen--Zener model with $H_{diag}=const$ and $%
H_{nond}=V_{0}\sec h(rt)$, the Demkov--Kunike model with $%
H_{diag}=E+E_{0}\tanh (rt)$ and $H_{nond}=V_{0}\sec h(rt)$ or $%
H_{nond}=const $, the Landau--Zener model with $H_{diag}=\lambda t$ and $%
H_{nond}=const$, and the parabolic model with $H_{diag}=at^{2}+b$ and $%
H_{nond}=const$. Although the last model has not an exact solution, it has
been studied comprehensively (see, for example, Ref. \cite{Suominen92}).
Here, $H_{diag}$ is the amplitude of the diagonal component of the
Hamiltonian (i.e., the two-level splitting) and $H_{nond}$ is the amplitude
of the nondiagonal component (i.e., the coupling perturbation). The symbols $%
V_{0}$, $E$, $E_{0} $, $r$ can represent arbitrary constants and $t$ is
time. One can find these models and their solutions in, for example, \cite%
{Suominen95}.

The advantage of these models comes from the possibility to reduce their Schr%
\"{o}dinger equations to second order differential equations (with time
dependent coefficients) whose solutions are known. For example, the Schr\"{o}%
dinger equation for the Rosen-Zener model can be reduced to an equation
whose solution is the hypergeometric function and the Schr\"{o}dinger
equation for the Landau-Zener model can be solved in terms of parabolic
cylinder functions. For an arbitrary Hamiltonian, the quasi-classical
approach based on the Wentzel-Kramers-Brillouin (WKB) approximation was
developed by Dykhne \cite{Dykhne61-62}--\cite{DykhneChaplik}. In this
method, the amplitude of the nonadiabatic transition is calculated under the
assumption that the adiabatic levels cross in the complex time plane and the
main contribution of the nonadiabatic coupling is given at the time of the
level crossing. The latter is usually the branch point of the phase integral
of the energies. The result becomes universal, i.e., independent of the
nonadiabatic coupling model and it is sufficient to take the leading term of
the coupling at the crossing point.

In this paper we consider adiabatic following of the dark, nonabsorbing
state in the three-level system excited by two resonant fields. The
adiabatic following of the dark state results in STIRAP (population
transfer). It was proposed in Ref. \cite{OregHioeEber84} where the so called
counterintuitive Raman pulse sequence emerged as a result of the search for
the generalization of the Liouville equations for the $N$-level system using
$SU(N)$ coherence vector theory \cite{HioeEber81}--\cite{OregHioeEber84}.
This pulse sequence consists of two fields coupling two low energy levels $1$
and $2$ with one common excited state $3$. If one of the levels (for
example, $2$) is initially empty, then applying first the field which
couples this state with level $3$, and then the field coupling the populated
state (for example, $1$) with $3$, it is possible to transfer the population
of state $1$ to state $2$, without appreciably populating the intermediate
state $3$.

Later, the importance of the adiabaticity in the counterintuitive pulse
sequence development was realized in \cite{KuklGaubHioeBergm89}--\cite%
{CarrHioe90}. There is a particular superposition of states $1$ and $2$ that
does not interact with the coupling fields and, if the development of the
field amplitudes in time is properly chosen, this superposition state
changes from state $1$ to state $2$. If the three-level atom follows this
superposition state, the atom population is transferred from state $1$ to
state $2$ by the pulse sequence without populating the intermediate excited
state $3$. This noncoupled superposition state was first introduced by
Arimondo \cite{Arimondo96} to explain qualitatively the dark resonance as
population trapping in this state \cite{ArimOrr76}--\cite{Orriol79}. The
noncoupled state is often referred to as a dark state.

It is obvious that, if the dark state changes in time, a process must exist
which tends to empty this state. The condition minimizing the dark state
depopulation is formulated in \cite{KuklGaubHioeBergm89}. This is done {\it %
ad hoc}, without an estimation of the nonadiabatic correction for the
excited probability amplitude. However, numerical calculations show that, if
this condition is satisfied, the adiabatic population transfer $1\rightarrow
2$ is almost perfect. Some attempts were undertaken to find a rigorous
justification of the intuitively found adiabaticity condition and to
estimate the amplitude of the excited state during the STIRAP pulse sequence
(see, for example, Refs. \cite{Stenholm96}--\cite{Fleischh99}). In \cite%
{Stenholm96}, the so called ramp pulses were considered, which allow an
exact solution. Furthermore, with the help of the method developed for the
two-level system by Dykhne (see the discussion above and Refs. \cite%
{Pechukas76}--\cite{Pechukas77}), the nonadiabatic amplitude of the excited
state $3$ is estimated for Gaussian and hyperbolic secant pulses. This is
possible because the Liouville equation for the two-level atom in terms of
the $SU(2)$ coherence vector (Bloch-vector model, see, for example, \cite%
{AllenEber75}) coincides with the Schr\"{o}dinger equation for the state
probability amplitudes of the three-level atom excited by two resonant
fields (see, for example, \cite{CarrHioe90}, \cite{Stenholm96}, \cite%
{Shakhmuratov2001}). However, the authors of \cite{Stenholm96} admit that
the analytical approximations for the nonadiabatic corrections ''{\it have
been introduced ad hoc without derivation}'' and they ''{\it would really
like to see more detailed investigations of the analytic behavior of the
system discussed}''.

Fleischhauer with coauthors \cite{Fleischh99} developed a different approach
introducing higher-order trapping states. They define a $n$th--order
generalized adiabatic basis, which is similar to the superadiabatic basis
introduced for the two-level system in \cite{Berry87}--\cite{Lim93}. By
successive transformations from the diabatic basis to the adiabatic basis,
then from this adiabatic basis, which can be considered as the adiabatic
basis of the first order, to the next, i.e., the second order adiabatic
basis, etc., the solution is presented as an infinite product of
transformations. The general expression for the $n$-th transformation matrix
is presented. However, it is hard to implement this scheme for an arbitrary
pulse sequence.

In this paper we develop a new method that allows us to calculate the
adiabatic and nonadiabatic components of the solution with controllable
accuracy. The adiabatic component describes the part of the atomic
probability amplitude visiting the excited atomic state during the pulse
train and coming back to the dark state when the pulses are gone. The atom
evolution following the adiabatic component of the solution resembles the
excitation--de-excitation process induced by a soliton in the two-level
atom. The nonadiabatic component is that part which is lost from the dark
state and describes the fraction of the atomic probability amplitude that is
left excited after the pulse train. We compare our result with the previous
theories reported in \cite{Stenholm96} and \cite{Fleischh99}. We found
corrections to the theory presented in \cite{Stenholm96} and show that our
result corresponds to the calculation of the infinite number of
transformations proposed in \cite{Fleischh99}. As an example of the validity
of the method, we present also the solution of the Rosen--Zener and
Landau-Zener models by our method (see the Appendix and the end of Sections
VI and VII).

The paper is organized as follows. In Sec. II we present the general
formalism employed in the description of the three-level atom excited by two
resonant fields. The transformation to the basis of the bright and dark
states is derived. It is shown that the system evolves between two states,
i.e., ''bright'' and ''common'' (they are specified in Section II). In Sec.
III we consider the time evolution of the atomic state vector for the $%
\Lambda $-scheme of excitation if the field amplitudes are time dependent.
Bloch-like equations are derived. In Sec. IV the adiabatic following
approximation is presented. The adiabatic solution for the stimulated Raman
adiabatic passage (STIRAP) is found in Sec. V. Nonadiabatic corrections are
described in Sec. VI. The case when the Raman Rabi frequency changes in time
is considered in Sec. VII-IX.

\section{THREE-LEVEL ATOM INTERACTING WITH TWO RESONANT FIELDS. GENERAL
FORMALISM}

We consider a three-level atom shown in Fig. 1 (a). The arrows indicate the
transitions induced by the coherent fields. This excitation scheme is
classified as the $\Lambda $ scheme. We define the level that is common for
both transitions as $3$. The others are designated by the numbers $1$ and $2$%
, level $2$ being of higher energy than $1$ and initially not populated. The
dynamic part of the Hamiltonian of this three-level atom, excited by two
resonant fields ${\bf E}_{1}(t)={\bf E}_{1}\cos (\Omega _{1}t+\varphi _{1})$
and ${\bf E}_{2}(t)={\bf E}_{2}\cos (\Omega _{2}t+\varphi _{2})$, is
\begin{equation}
H=\sum_{n=1}^{3}\omega _{n}\widehat{P}_{nn}-\left( B_{1}\widehat{P}%
_{13}e^{i\Omega _{1}t+i\varphi _{1}}+B_{2}\widehat{P}_{23}e^{i\Omega
_{2}t+i\varphi _{2}}+h.c.\right) ,  \label{Eq1}
\end{equation}%
where $\omega _{n}$ is the energy of the state $n$. Planck's constant is set
equal one ($\hbar =1$) for simplicity. The operators $\widehat{P}_{mn}$ are
defined by $\widehat{P}_{mn}=\left| m\right\rangle \left\langle n\right| $,
where $\left\langle n\right| $ and $\left| m\right\rangle $ are bra and ket
vectors of the states $n$ and $m$ in the Schr\"{o}dinger representation. The
interaction constant (Rabi frequency) $B_{n}=\left( {\bf d}_{n3}\cdot {\bf E}%
_{n}\right) /2=\left( {\bf d}_{3n}\cdot {\bf E}_{n}\right) /2$ depends on
the dipole-transition matrix element, taken real so, ${\bf d}_{n3}={\bf d}%
_{3n}$, and on the field amplitude ${\bf E}_{n}$. The rotating wave
approximation is taken into account.

If the fields ${\bf E}_{1}(t)$ and ${\bf E}_{2}(t)$ are in exact resonance
with the relevant transitions, the Hamiltonian, Eq. (\ref{Eq1}), can be made
slowly varying by transforming it into the interaction representation (IR).
This representation is defined by a canonical transformation by means of the
unitary operator
\begin{equation}
T=\exp \left( i\sum_{n=1}^{3}\omega _{n}\widehat{P}_{nn}t\right) .
\label{Eq2}
\end{equation}%
The wave function of the atom in the IR is defined by $\left| \Phi
(t)\right\rangle =T\left| \Psi (t)\right\rangle $, where $\left| \Psi
(t)\right\rangle $ is the wave function in the Schr\"{o}dinger
representation. $\left| \Phi (t)\right\rangle $ satisfies the Schr\"{o}%
dinger equation with the effective Hamiltonian \cite{Abragam61}
\begin{equation}
{\cal H}=THT^{-1}+i\stackrel{.}{T}T^{-1}.  \label{Eq3}
\end{equation}%
This Hamiltonian has the explicit form
\begin{equation}
{\cal H}=-B_{1}\widehat{P}_{13}e^{i\varphi _{1}}-B_{2}\widehat{P}%
_{23}e^{i\varphi _{2}}+h.c.,  \label{Eq4}
\end{equation}%
or in matrix notation
\begin{equation}
{\cal H}=-\left[
\begin{array}{lll}
\;0 & \;0 & B_{1}e^{i\varphi _{1}} \\
\;0 & \;0 & B_{2}e^{i\varphi _{2}} \\
B_{1}e^{-i\varphi _{1}} & B_{2}e^{-i\varphi _{2}} & \;0%
\end{array}%
\right] .  \label{Eq5}
\end{equation}%
The $\widehat{P}_{MN}$-operators are defined for the vectors $\left|
M\right\rangle $ and $\left\langle N\right| $ of the interaction
representation differing from the states $\left| m\right\rangle $ and $%
\left\langle n\right| $ by the phase factors $\exp (-i\omega _{m}t)$ and $%
\exp (i\omega _{n}t)$.

Assume that the field amplitudes $B_{1}$, $B_{2}$ and the phases $\varphi
_{1}$, $\varphi _{2}$ are constant. Then the Hamiltonian, Eqs. (\ref{Eq4})--(%
\ref{Eq5}), is diagonalized by a unitary transformation
\begin{equation}
\overline{{\cal H}}=Q{\cal H}Q^{-1}=\sqrt{B_{1}^{2}+B_{2}^{2}}\left(
\widehat{P}_{\overline{3}\overline{3}}-\widehat{P}_{\overline{1}\overline{1}%
}\right) ,  \label{Eq6}
\end{equation}%
where
\begin{equation}
Q=\frac{1}{\sqrt{2}}\left[
\begin{array}{lll}
\;\;\ e^{-i\varphi _{1}}\sin \alpha & \;\;\ \ e^{-i\varphi _{2}}\cos \alpha
& \;\;\ \ 1 \\
\sqrt{2}e^{i\varphi _{2}}\cos \alpha & -\sqrt{2}e^{i\varphi _{1}}\sin \alpha
& \;\ \ \ 0 \\
\;\;\;e^{-i\varphi _{1}}\sin \alpha & \;\ \ \ e^{-i\varphi _{2}}\cos \alpha
& \,-1%
\end{array}%
\right] .  \label{Eq7}
\end{equation}%
and $\tan \alpha =B_{1}/B_{2}$. The new ket vectors $\left| \overline{n}%
\right\rangle $ are related to the former ones $\left| N\right\rangle $,
defined in the interaction representation, as follows $\left| \overline{\Phi
}\right\rangle =Q\left| \Phi \right\rangle $. The explicit relations are
\begin{equation}
\left| \overline{1}\right\rangle =\frac{\sqrt{2}}{2}\left( e^{i\varphi
_{1}}\sin \alpha \,\left| 1\right\rangle +e^{i\varphi _{2}}\cos \alpha
\,\left| 2\right\rangle +\left| 3\right\rangle \right) ,  \label{Eq8}
\end{equation}%
\begin{equation}
\left| \overline{2}\right\rangle =e^{-i\varphi _{2}}\cos \alpha \,\left|
1\right\rangle -e^{-i\varphi _{1}}\sin \alpha \,\left| 2\right\rangle ,
\label{Eq9}
\end{equation}%
\begin{equation}
\left| \overline{3}\right\rangle =\frac{\sqrt{2}}{2}\left( e^{i\varphi
_{1}}\sin \alpha \,\left| 1\right\rangle +e^{i\varphi _{2}}\cos \alpha
\,\left| 2\right\rangle -\left| 3\right\rangle \right) .  \label{Eq10}
\end{equation}%
The basis $\left| \overline{n}\right\rangle $, in which the Hamiltonian is
diagonal, is called the basis of the quasi-energy states \cite{Zeld73}, \cite%
{Shakhmuratov77}. It coincides with the basis of the dressed states if the
limit of the classical field is taken for them.

States $\left| \overline{1}\right\rangle $ and $\left| \overline{3}%
\right\rangle $ are mixtures of all unperturbed states $\left|
1\right\rangle $, $\left| 2\right\rangle $ and $\left| 3\right\rangle $,
whereas state $\left| \overline{2}\right\rangle $ does not contain the
common state $\left| 3\right\rangle $. If the atom is in state $\left|
\overline{2}\right\rangle $, a mixture of the ground state sublevels $1$ and
$2$, the atom does not leave this state since it is an eigenstate of the
interaction Hamiltonian, Eq. (\ref{Eq6}). The bichromatic field ${\bf E}(t)=%
{\bf E}_{1}(t)+{\bf E}_{2}(t)$ does not interact with such an atom, and,
consequently, it is not excited. Therefore, we call this state a dark state
and designate it $\left| d\right\rangle $. This state was introduced for the
first time by Arimondo in Ref. \cite{Arimondo96}, who called this state a
non-coupled state. He introduced the coupled state as well, which interacts
with the bichromatic field ${\bf E}(t)$. Following Arimondo we define the
state
\begin{equation}
\left| b\right\rangle =e^{i\varphi _{1}}\sin \alpha \,\left| 1\right\rangle
+e^{i\varphi _{2}}\cos \alpha \,\left| 2\right\rangle ,  \label{Eq11}
\end{equation}%
which is orthogonal to the dark state. We call this state a bright state,
since for the $\Lambda $-scheme, if the atom is in this state, it is excited
by the bichromatic field and then the luminescence from state $\left|
3\right\rangle $ may follow. Excitation can take place because the bright
state is not an eigenstate of the interaction Hamiltonian, Eq. (\ref{Eq6}).

The states $\left| d\right\rangle $, $\left| b\right\rangle $ and $\left|
3\right\rangle $ are mutually orthogonal and can be chosen as a new basis.
We call this basis $dbc$, designating the state $\left| 3\right\rangle $ by
the letter $\left| c\right\rangle $, since it is common for the induced
transitions. Because we will often refer to these states, they are presented
below in a common set of equations to simplify further citation
\begin{equation}
\left| d\right\rangle =e^{-i\varphi _{2}}\cos \alpha \,\left| 1\right\rangle
-e^{-i\varphi _{1}}\sin \alpha \,\left| 2\right\rangle ,  \label{Eq12}
\end{equation}%
\begin{equation}
\left| b\right\rangle =e^{i\varphi _{1}}\sin \alpha \,\left| 1\right\rangle
+e^{i\varphi _{2}}\cos \alpha \,\left| 2\right\rangle ,  \label{Eq13}
\end{equation}%
\begin{equation}
\left| c\right\rangle =\left| 3\right\rangle .  \label{Eq14}
\end{equation}%
The interaction Hamiltonian, Eqs. (\ref{Eq4})--(\ref{Eq5}), is transformed
in this basis as follows
\begin{equation}
{\cal H}_{dbc}=S{\cal H}S^{-1}=-B\left( \widehat{P}_{bc}+\widehat{P}%
_{cb}\right) ,  \label{Eq15}
\end{equation}%
or in a matrix form
\begin{equation}
{\cal H}_{dbc}=-\left[
\begin{array}{lll}
0 & 0 & 0 \\
0 & 0 & B \\
0 & B & 0%
\end{array}%
\right] ,  \label{Eq16}
\end{equation}%
where $B=\sqrt{B_{1}^{2}+B_{2}^{2}}$ and the unitary operator of the
canonical transformation is
\begin{equation}
S=\left[
\begin{array}{ccc}
e^{i\varphi _{2}}\cos \alpha & \;-e^{i\varphi _{1}}\sin \alpha & \;\;0 \\
e^{-i\varphi _{1}}\sin \alpha & \;e^{-i\varphi _{2}}\cos \alpha & \;\;0 \\
0 & \;0 & \;\;1%
\end{array}%
\right] .  \label{Eq17}
\end{equation}%
We call this basis the $dbc$-representation.

Figure 1 (b) shows the excitation scheme in the $dbc$--basis where only the $%
b$ and $c$ states are coupled. The $b$ and $c$ states do not correspond to a
defined energy. To present schematically the $dbc$ states and the
transitions between them, we choose the state population before the
excitation for their relative position in the diagram. So, the vertical
scale in Fig. 1 (b) represents the initial population of the states counted
from the bottom to the top.

The Hamiltonian in the $dbc$-basis, Eq. (\ref{Eq15}), resembles the
interaction Hamiltonian of the two-level system $bc$, excited by one
resonant field with an effective interaction constant $B$. The dynamic
evolution of the atom is described by the Schr\"{o}dinger equation with this
Hamiltonian containing only the transition operators between states $b$ and $%
c$. If, for example, initially the $d$ and $b$ states are populated and $c$
is not,
\begin{equation}
\left| \Phi _{bc}(0)\right\rangle =C_{d}\left| d\right\rangle +C_{b}\left|
b\right\rangle ,  \label{Eq18}
\end{equation}%
the dynamics of the three-level atom follows the very simple formula
\begin{equation}
\left| \Phi _{bc}(t,\pm )\right\rangle =C_{b}\left[ \cos \left( \frac{\chi
_{R}t}{2}\right) \left| b\right\rangle +i\sin \left( \frac{\chi _{R}t}{2}%
\right) \left| c\right\rangle \right] +C_{d}\left| d\right\rangle ,
\label{Eq19}
\end{equation}%
where $\chi _{R}=2B$ is the Raman Rabi frequency. This Rabi frequency is
defined as twice the interaction constant in the Hamiltonian, Eq. (\ref{Eq15}%
), according to the conventional definition adopted for the two-level system %
\cite{AllenEber75}. The initial condition specified in Eq. (\ref{Eq18}) has
a simple relation to the IR basis
\begin{equation}
\left| \Phi _{bc}(0)\right\rangle =(C_{b}e^{i\varphi _{1}}\sin \alpha
+C_{d}e^{-i\varphi _{2}}\cos \alpha )\left| 1\right\rangle
+(C_{b}e^{i\varphi _{2}}\cos \alpha -C_{d}e^{-i\varphi _{1}}\sin \alpha
)\left| 2\right\rangle .  \label{Eq20}
\end{equation}%
If the initial condition is $\left| \Phi _{bc}(0)\right\rangle =\left|
1\right\rangle $, then $C_{b}=e^{-i\varphi _{1}}\sin \alpha $ and $%
C_{d}=e^{i\varphi _{2}}\cos \alpha $, which means that the phase relation
between the amplitudes of $b$ and $d$ states\ is fixed as $%
C_{b}=C_{d}e^{-i(\varphi _{1}-\varphi _{2})}\tan \alpha $.

Suppose that the pulse duration of the bichromatic field, $t_{2\pi }$, is
chosen such that the effective two-level system $bc$ experiences a so called
$2\pi $-pulse, i.e., $\chi _{R}t_{2\pi }=2\pi $. During the first half of
the pulse, the population of level $b$ is completely transferred to state $c$%
. During the second half of the pulse, the population of the excited state $%
c $ is transferred back to state $b$. The state vector of the atom at the
end of the pulse is
\begin{equation}
\left| \Phi _{bc}(t_{2\pi })\right\rangle =-C_{b}\left| b\right\rangle
+C_{d}\left| d\right\rangle .  \label{Eq21}
\end{equation}%
Equation (\ref{Eq21}) shows that the $2\pi $-pulse results in a $\pi $ phase
shift of the $b$ state probability amplitude. Usually, for the simple
two-level atom, this $\pi $ phase shift of the ground state after a $2\pi $%
-pulse is not revealed in the observables. In our case, apart from the two
levels $b$ and $c$, we have the third level $d$ whose phase is well defined
with respect to the phase of state $b$. The phase definition is determined
by the phase relation of the components of the bichromatic field [see
comments after Eq. (\ref{Eq20})]. In section IV we show the importance of
this point.

\section{TIME DEPENDENT AMPLITUDES}

If during the pulse excitation the field amplitudes ${\bf E}_{1}$, ${\bf E}%
_{2}$ are time dependent and they have the same time evolution satisfying
the condition $B_{1}(t)/B_{2}(t)=\tan \alpha =const$, then the developments
of the $\left| d\right\rangle $, $\left| b\right\rangle $ states in the
ket-vectors $\left| 1\right\rangle $, $\left| 2\right\rangle $ have constant
coefficients and the $S$-transformation is time independent. In this case
the solution of the Schr\"{o}dinger equation is simple and the development
coefficients of the state vector $\left| \Phi _{dbc}\right\rangle $,
\begin{equation}
\left| \Phi _{dbc}(t)\right\rangle =C_{d}\left| d\right\rangle +C_{b}\left|
b\right\rangle +C_{c}\left| c\right\rangle ,  \label{Eq22}
\end{equation}%
are $C_{d}(t)=C_{d}(0)$, $C_{b}(t)=C_{b}(0)\cos [\theta (t)/2]$ and $%
C_{c}(t)=iC_{b}(0)\sin [\theta (t)/2]$, where $C_{d}(0)$, $C_{b}(0)$ are the
initial values of the probability amplitudes and $\theta (t)=2\int_{-\infty
}^{t}B(\tau )d\tau $ is the pulse area of the bichromatic field. This case
corresponds to so called matched pulses \cite{Harris94}--\cite{Harris95}.

If there is a time shift $T$ between the pulses ${\bf E}_{1}(t)$ and ${\bf E}%
_{2}(t)$, one can again introduce the bright and dark states, employing the
time dependent $S$-transformation. Then the evolution of the state vector of
the atom in the $dbc$-basis is given by
\begin{equation}
\left| \Phi _{dbc}(t)\right\rangle =S(t)\left| \Phi (t)\right\rangle ,
\label{Eq23}
\end{equation}%
where $\left| \Phi (t)\right\rangle $ is the state vector in the interaction
representation. Taking the time derivative of the equation (\ref{Eq23}), one
can obtain the Schr\"{o}dinger equation
\begin{equation}
\frac{d\left| \Phi _{dbc}\right\rangle }{dt}=-i\overline{{\cal H}}%
_{dbc}\left| \Phi _{dbc}\right\rangle  \label{Eq24}
\end{equation}%
with the modified Hamiltonian
\begin{equation}
\overline{{\cal H}}_{dbc}={\cal H}_{dbc}+i\stackrel{.}{S}S^{-1},
\label{Eq25}
\end{equation}%
where ${\cal H}_{dbc}$ is defined in Eq. (\ref{Eq15}) and
\begin{equation}
i\stackrel{.}{S}S^{-1}=i\stackrel{.}{\alpha }\left[ \widehat{P}%
_{bd}e^{-i\left( \varphi _{1}+\varphi _{2}\right) }-\widehat{P}%
_{db}e^{i\left( \varphi _{1}+\varphi _{2}\right) }\right]  \label{Eq26}
\end{equation}%
The first part, ${\cal H}_{dbc}$, of the modified Hamiltonian induces
transitions between the $\left| b\right\rangle $ and $\left| c\right\rangle $
states with the rate $\chi _{R}=2B$, and the second part, $i\stackrel{.}{S}%
S^{-1}$, induces transitions between states $\left| d\right\rangle $ and $%
\left| b\right\rangle $ with the rate $2\stackrel{.}{\alpha }$ (see Fig. 2).

The development coefficients of the state vector $\left| \Phi
_{dbc}\right\rangle $ satisfy the equations
\begin{equation}
\stackrel{.}{Z}_{d}=-\stackrel{.}{\alpha }Y_{b}  \label{Eq27}
\end{equation}%
\begin{equation}
\stackrel{.}{Y}_{b}=-BX_{c}+\stackrel{.}{\alpha }Z_{d}  \label{Eq28}
\end{equation}%
\begin{equation}
\stackrel{.}{X}_{c}=BY_{b}  \label{Eq29}
\end{equation}%
where the substitution $Z_{d}=C_{d}\exp (-i\varphi _{2})$, $Y_{b}=C_{b}\exp
(i\varphi _{1})$, $X_{c}=-iC_{c}\exp (i\varphi _{1})$ is made to deal with
real numbers. Here the phases $\varphi _{1}$ and $\varphi _{2}$ are assumed
to be constant throughout the excitation process. Equations (\ref{Eq27})--(%
\ref{Eq29}) remarkably coincide with the Bloch equations for an abstract
two-level system $g-e$ excited by a field with frequency $\omega $ slightly
tuned from resonance ($g$ and $e$ being ground and excited states,
respectively) \cite{Shakhmuratov2001}. Expressed in terms of the
Bloch-vector components, which are the following combinations of the density
matrix of the two-level system $\rho $: $u-iv=2\rho _{ge}\exp (-i\omega t)$
and $w=\rho _{gg}-\rho _{ee}$, these equations are, Ref. \cite{AllenEber75},
\begin{equation}
\stackrel{.}{w}=-\chi v,  \label{Eq30}
\end{equation}%
\begin{equation}
\stackrel{.}{v}=-\Delta u+\chi w,  \label{Eq31}
\end{equation}%
\begin{equation}
\stackrel{.}{u}=\Delta v.  \label{Eq32}
\end{equation}%
Here $\Delta $ is the detuning from resonance and $\chi $ is the Rabi
frequency. If one makes the substitution $w=Z_{d}$, $v=Y_{b}$, $u=X_{c}$ for
the variables and $\chi =\stackrel{.}{\alpha }$, $\Delta =B$ for the
parameters, both sets of equations are identical.

The coincidence of the equations enables us to use the adiabatic following
approach developed by Crisp \cite{Crisp1973} to describe nonresonant
excitation of the two-level system. A similar approach was applied by Laine
and Stenholm (LS) \cite{Stenholm96} based on the ideas of the adiabatic
following developed by Dykhne \cite{Dykhne1970},\cite{DykhneChaplik} for the
two-level system with time dependent splitting and coupling parameters. In
the LS approach,\cite{Stenholm96}, the instantaneous eigenstates of the
three-level atom excited by two resonant pulses are found. This basis is
called the adiabatic representation. The transformation to the instantaneous
basis [see Eq. (\ref{Eq7})] and the instantaneous Hamiltonian diagonal in
this basis [see Eq. (\ref{Eq6})] are time dependent. Therefore, the
adiabatic states coupling, $i\stackrel{.}{Q}Q^{-1}$, also appears in their
consideration. In spite of being different in structure, our Schr\"{o}dinger
equation in the changing $dbc$-basis and the LS equation in the
instantaneous eigenstate basis can both be reduced to the equation for the
two-level system. However, our equations (\ref{Eq27})--(\ref{Eq29}) are in a
one to one corespondence to the Bloch equations while the LS equations match
the equations for the two-level density matrix elements $\rho _{eg}$, $\rho
_{ge}$ and $\rho _{gg}-\rho _{ee}$. Although this difference is not crucial,
it brings, however, some convenience in our case, because our equations are
expressed for real quantities.

\section{ADIABATIC FOLLOWING APPROXIMATION}

Crisp gave in his paper \cite{Crisp1973} a qualitative and quantitative
description of the adiabatic following approximation for the two-level atom
excited by a nonresonant field. The qualitative explanation employs the
Bloch-vector model as follows. The Bloch equations can be written in the
form
\begin{equation}
\stackrel{.}{{\bf R}}={\bf H}_{eff}\times {\bf R},  \label{Eq33}
\end{equation}%
if the relaxation times $T_{1}$ and $T_{2}$ are very long compared with the
pulse duration. The Bloch vector ${\bf R}$ is given by%
\begin{equation}
{\bf R}=X{\bf e}_{x}+Y{\bf e}_{y}+Z{\bf e}_{z},  \label{Eq34}
\end{equation}%
and the effective field ${\bf H}_{eff}$ is
\begin{equation}
{\bf H}_{eff}=-\chi {\bf e}_{x}+\Delta {\bf e}_{z}.  \label{Eq35}
\end{equation}%
Equation (\ref{Eq33}) has the geometric interpretation that the Bloch vector
tries to precess about the effective field as ${\bf H}_{eff}$ varies both in
magnitude and direction. If the condition $\left| \Delta \right| \gg \left|
\chi \right| $ is satisfied, the precession frequency of the Bloch-vector, $%
\sqrt{\chi ^{2}+\Delta ^{2}}$, will be large compared with the rate of
change of the field $\chi (t)$. If initially the Bloch-vector points
parallel or antiparallel to ${\bf e}_{z}$ [$X(0)=Y(0)=0$ and $Z(0)=\pm 1$],
then precessing with high frequency $\left| \Delta \right| $ about the
effective field, the Bloch-vector remains nearly parallel or antiparallel to
the effective field as it moves adiabatically. Qualitatively, adiabatic
movement of the vector ${\bf H}_{eff}$ means that the Bloch-vector does not
slip off from the direction of the effective field if the rotation of the
Bloch-vector around the ${\bf H}_{eff}$--field is much faster than the
change of the ${\bf H}_{eff}$--field direction. Quantitatively, this sets a
condition that the precession period $T_{p}\simeq 2\pi /\left| \Delta
\right| $ is to be much shorter than the time scale of the effective field
variation. In this limit, the angle between the Bloch vector and the
effective field will remain very small. The deviation of the Bloch vector
from the effective field is described by the nonadiabatic corrections. In
the next section we consider how small this deviation angle is for
particular pulses.

The approximation of the adiabatic following allows also a quantum
mechanical interpretation. The Hamiltonian in the $dbc$ time varying
representation, Eq. (\ref{Eq25}), consists of two parts. One part couples
two, initially unpopulated states, $b$ and $c$. The coupling parameter $B$
is time varying. The other part couples the populated state $d$ with the
unpopulated state $b$. The coupling parameter $\stackrel{.}{\alpha }$ is
also time varying. In order that the atom is subject to the adiabatic
following of the changing $d$ state (the atom stays in this state), one has
to choose the pulse train with the Raman Rabi frequency $B$ much larger than
$\stackrel{.}{\alpha }$ when the latter takes its maximum value \cite%
{Shakhmuratov2001}. Also the time variation scale of $\stackrel{.}{\alpha }$
must be much longer than the Raman Rabi period. These conditions can be made
plausible by the following speculations.

Since initially only state $d$ is populated, the atom starts to evolve when $%
\stackrel{.}{\alpha }(t)$ becomes nonzero. When the slow process $\stackrel{.%
}{\alpha }$ brings a small fraction of the population to state $b$, the fast
Rabi oscillation via the $2\pi $-cycle (see the discussion at the end of
Section II) changes the sign of the $Y_{b}$-amplitude from plus to minus,
reversing back the population transfer process from $d\rightarrow b$ to $%
b\rightarrow d$. As a result, the population of the $b$-state does not
increase and remains small. One can roughly estimate its mean value by
solving the equations (\ref{Eq27})--(\ref{Eq29}) with the initial condition $%
X_{c}(0)=Y_{b}(t)=0$, $Z_{d}(0)=1$ and assuming that $\chi _{R}$ and $%
\stackrel{.}{\alpha }$ are constant, i.e., neglecting the derivatives of
these parameters,
\begin{equation}
Z_{d}(t)=\frac{B^{2}}{B^{2}+\stackrel{.}{\alpha }^{2}}+\frac{\stackrel{.}{%
\alpha }^{2}}{B^{2}+\stackrel{.}{\alpha }^{2}}\cos \sqrt{B^{2}+\stackrel{.}{%
\alpha }^{2}}t,  \label{Eq36}
\end{equation}%
\begin{equation}
Y_{b}(t)=\frac{\stackrel{.}{\alpha }}{\sqrt{B^{2}+\stackrel{.}{\alpha }^{2}}}%
\sin \sqrt{B^{2}+\stackrel{.}{\alpha }^{2}}t,  \label{Eq37}
\end{equation}%
\begin{equation}
X_{c}(t)=\frac{\stackrel{.}{\alpha }B}{B^{2}+\stackrel{.}{\alpha }^{2}}%
\left( 1-\cos \sqrt{B^{2}+\stackrel{.}{\alpha }^{2}}t\right) .  \label{Eq38}
\end{equation}%
The solution shows that the change of the amplitude of the $d$-state is
small if $B=\sqrt{B_{1}^{2}+B_{2}^{2}}\gg \,\stackrel{.}{\alpha }$. This is
exactly the condition of the adiabatic following formulated in Ref. \cite%
{KuklGaubHioeBergm89} for STIRAP.

It should be emphasized that the higher time derivatives of the mixing
parameter $\alpha (t)$, neglected above, play a crucial role in the
adiabatic population transfer. However, the solution (\ref{Eq36})--(\ref%
{Eq38}) gives some qualitative hint on the choice of the relation between $B$
and $\stackrel{.}{\alpha }$.

\section{ADIABATIC SOLUTION}

The adiabatic following approximation can be applied to the consideration of
STIRAP and electromagnetically induced transparency (EIT) because in both
cases the atom follows the dark, noncoupled state. A first attempt to study
adiabatic following in EIT was undertaken in Ref. \cite{Shakhmuratov2001}.
In this section we consider the application of this approach to STIRAP.

The stimulated Raman adiabatic passage assumes that before the application
of the $E_{1}(t)$ and $E_{2}(t)$ pulses the atom is in state $\left|
1\right\rangle $. The duration of the excitation as well as the pulse
sequence must be chosen such that at the end of the pulse train the atom is
left in state $\left| 2\right\rangle $. It is expected that during this
process the atom stays in the dark state. However, this state itself changes
since the coefficients $\cos \alpha $ and $\sin \alpha $ of the development
of the dark state in the vectors $\left| 1\right\rangle $ and $\left|
2\right\rangle $ [see Eq. (\ref{Eq12})] change in time. The parameter $%
\alpha $ rises from zero to $\pi /2$, so $\left| d\right\rangle =\left|
1\right\rangle $ before the excitation and $\left| d\right\rangle
=-e^{-i(\varphi _{1}-\varphi _{2})}\left| 2\right\rangle $ after it. Since $%
\tan \alpha =B_{1}/B_{2}$, the condition imposed on $\alpha $ means that the
$B_{1}(t)$-pulse must be delayed with respect to the $B_{2}(t)$-pulse. We
have to emphasize that the phase of the final state $2$ after the pulse
train is related to the phase of the initial state $1$ according to Eq. (\ref%
{Eq12}).

If we choose two identical, bell-shaped, delayed pulses having a hyperbolic
secant shape, then the interaction constants evolve in time as follows
\begin{equation}
B_{n}(t)=B_{0}\sec h[r(t-t_{n})],  \label{Eq39}
\end{equation}%
where $n=1$ or $2$, $t_{n}$ is the time when the $n$-pulse has maximum
amplitude, and $r$ is the rise and fall rate of the pulse edges. The mixing
parameter $\alpha (t)$ increases monotonously if the condition $t_{1}>t_{2}$
is satisfied. This pulse sequence was considered by Laine and Stenholm in
Ref. \cite{Stenholm96}.

Let us analyze the constraints imposed on the parameters $r$, $t_{1}$ and $%
t_{2}$ to have $\alpha $ changed from zero to $\pi /2$. The mixing angle $%
\alpha $ varies according to the relation $\tan \alpha (t)=B_{1}(t)/B_{2}(t)$%
. For the pulse sequence specified above, an explicit form of this relation
is
\begin{equation}
\tan \alpha (t)=\frac{1+D\tanh [r(t-t_{0})]}{1-D\tanh [r(t-t_{0})]},
\label{Eq40}
\end{equation}%
where $D=\tanh (rT/2)$, $t_{0}=\left( t_{1}+t_{2}\right) /2$ is the mean
time of the maxima and $T=t_{1}-t_{2}$ is the time interval between the
maxima of the pulses. Suppose that, initially, the atom is in the ground
state $1$ and we start the atom evolution from the initial mixing parameter $%
\alpha _{in}(-\infty )$ satisfying the condition $\tan (\alpha _{in})=0.01$,
which means that essentially $C_{d}(-\infty )e^{-i\varphi _{2}}=\cos (\alpha
_{in})\approx 1$ and $C_{b}(-\infty )e^{i\varphi _{1}}=\sin (\alpha
_{in})=10^{-2}$. We stop the atom evolution at $\tan [\alpha _{fin}(+\infty
)]=100$. So, if the final state coincides with state $2$ then $C_{d}(+\infty
)e^{-i\varphi _{1}}=\sin (\alpha _{fin})\approx 1$ and $C_{b}(-\infty
)e^{i\varphi _{2}}=-\cos (\alpha _{fin})=-10^{-2}$, where the phase change
of state $2$ is taken into account. From Eq. (\ref{Eq40}) it follows that $%
\tan \alpha (\pm \infty )=(1\pm D)/(1\mp D)$ and at the condition imposed on
$\alpha _{in}$ and $\alpha _{fin}$ we have $rT=4.6$. Since state $b$ is
strongly coupled with state $c$ [the coupling is $B(t)$], we have to keep
the initial population of state $b$ as small as possible [$C_{b}(-\infty )$
must be close to zero]. Otherwise, the probability amplitude $C_{b}(-\infty
) $, if not infinitely small, will be spread among the $b$ and $c$ states by
the pulse train and population transfer $1\rightarrow 2$ will be imperfect.
Therefore, on the one hand, the initial value of the mixing parameter $%
\alpha $ must be as small as possible to have complete population transfer $%
1\rightarrow 2$. For a small initial mixing angle $\alpha _{in}$, the
relation between $\alpha _{in}$ and $T$ becomes simple, i.e., $\tanh
(rT/2)\approx 1-2\alpha _{in}$. The smaller the initial value of the mixing
angle $\alpha $, the larger the product $rT$ or the pulse separation $T$ is.
On the other hand, if the distance between the pulses is large, the value of
the coupling $B(t)$ at $t_{0}$ becomes small: the larger the distance, the
smaller the coupling. However, the adiabatic following demands a large
coupling $B$ at $t_{0}$. Therefore, one has to choose the optimum value of
the pulse spacing satisfying two conditions simultaneously. The distance
between pulses is to be as large as possible to have a small value of $%
\alpha _{in}$ and, at the same time, $B(t_{0})$ should be kept as large as
possible. Below we give some arguments how to find this optimum value.

If $rT\rightarrow 0$, then according to Eq. (\ref{Eq40}) we have $\alpha
\rightarrow \pi /4$ throughout the excitation and the dark state does not
change in time. It has the probability amplitude $\left| C_{d}\right| =\sqrt{%
2}/2$. State $b$, having the initial population $\left| C_{b}(-\infty
)\right| ^{2}=1/2$, is depopulated with the rate $B$ [$C_{b}(t)=C_{b}(-%
\infty )\cos \left( \theta (t)/2\right) $, $C_{c}(t)=iC_{b}(-\infty )\sin
\left( \theta (t)/2\right) $, see the definition of $\theta (t)$ in Section
III]. For small values of $rT$, the time interval of the $\alpha $ change is
small. For example, if $rT=0.5$, then $\cos (\alpha _{in})=0.855$, $\sin
(\alpha _{in})=0.519$ and $\cos (\alpha _{fin})=0.519$, $\sin (\alpha
_{fin})=0.855$, which corresponds to the change of $\alpha $ from $\alpha
_{in}=0.174\pi $ to $\alpha _{fin}=0.326\pi $ during the pulse train. In
this case only the fraction $(0.855)^{2}=0.731$ of the atomic population is
transferred to state $\left| 2\right\rangle $ if the atom adiabatically
follows the dark state. Another fraction $(0.515)^{2}=0.269$ of the atomic
population participates in the process of excitation and de-excitation
between states $b$ and $c$. This means that the population transfer via the
change of the amplitude of the dark state components is possible only if the
time interval between the pulses exceeds a certain value. We choose the
value $rT=5$ since in this case the initial population of state $b$ is $%
4.5\cdot 10^{-5}$ so that we neglect this population in our further
consideration.

Figure 3 (a) shows the pulse train with $rT=5$ along with the change of the
mixing parameter $\alpha $ during the excitation. On the same plot the
dependence of the derivative
\begin{equation}
\stackrel{.}{\alpha }(t)=\frac{rD\sec h^{2}[r(t-t_{0})]}{1+D^{2}\tanh ^{2}%
\left[ r\left( t-t_{0}\right) \right] }=\frac{r\sinh (rT)}{1+\cosh (rT)\cosh
[2r(t-t_{0})]}  \label{Eq41}
\end{equation}%
is shown. This derivative takes a maximum value $\stackrel{.}{\alpha }_{\max
}(t_{0})=rD$ at $t=t_{0}$. For example, if $rT=5$, then $\stackrel{.}{\alpha
}_{\max }(t_{0})=0.987r$. In Fig. 3 (b) we compare the time evolution of the
Rabi frequency $\chi _{R}/2=B(t)=\sqrt{B_{1}^{2}(t)+B_{2}^{2}(t)}$ with the
evolution of the derivative $\stackrel{.}{\alpha }(t)$.

The Rabi frequency determines the transition rate between states $b$ and $c$%
, whereas the derivative of the state mixing angle $\alpha $ specifies the
transition rate between states $d$ and $b$ [see Eqs. (\ref{Eq27})--(\ref%
{Eq29})]. Since at $t=t_{0}$ the parameter $\stackrel{.}{\alpha }(t)$ takes
its maximum value and $B(t)$ has its minimum, the adiabatic following
condition is $B(t_{0})\gg \stackrel{.}{\alpha }(t_{0})$ or, explicitly, $%
\sqrt{2}B_{0}\gg r\sinh \left( rT/2\right) $. Figure 4 (solid lines) shows
the time dependence of the amplitudes $X_{c}$ (plot a), $Y_{b}$ (plot b) and
$Z_{d}$ (plot c) obtained by the numerical solution of the equations (\ref%
{Eq27})--(\ref{Eq29}) for the pulse train with $t_{1}=2.5/r$, $t_{0}=0$, $%
t_{2}=-2.5/r$, $T=5/r$, $B_{0}=42.8r$. The relation between $B_{0}$ and $r$
corresponds to the ratio $B(t_{0})/\stackrel{.}{\alpha }(t)=10$. We see that
the deviation of the $Z_{d}$--amplitude from its initial value ($\sim 0.005$%
) and the maximum absolute value of the $Y_{b}$--amplitude ($\sim 0.01$) are
different from the values predicted by the simplified solution, Eqs. (\ref%
{Eq36})--(\ref{Eq38}). Only the maximum value of $X_{c}$ ($\sim 0.1$) fits
the value of the simplified solution.

To estimate the actual values of the probability amplitudes of the $dbc$
states during the excitation and find the borders within which the adiabatic
following of the dark state takes place, we follow the theory developed by
Crisp \cite{Crisp1973} for the case when the condition $B(t_{0})\gg \stackrel%
{.}{\alpha }(t)$ is well satisfied.

In this case we can use the expansion in a power series of the parameter $%
\stackrel{.}{\alpha }(t)$ for the solution of the equations (\ref{Eq27})-(%
\ref{Eq29}). Then the first term of the expansion is found by setting $%
Z_{d}(t)=1$ in the equations (\ref{Eq28}),(\ref{Eq29}) and then solving
them. The solution is%
\begin{equation}
Y_{b}(t)=\int\nolimits_{-\infty }^{t}\cos \left[ \int_{\tau }^{t}B(\tau
_{1})d\tau _{1}\right] \stackrel{.}{\alpha }(\tau )d\tau ,  \label{Eq42}
\end{equation}%
\begin{equation}
X_{c}(t)=\int\nolimits_{-\infty }^{t}\sin \left[ \int_{\tau }^{t}B(\tau
_{1})d\tau _{1}\right] \stackrel{.}{\alpha }(\tau )d\tau .  \label{Eq43}
\end{equation}%
This is the part of the general solution that is linear in $\stackrel{.}{%
\alpha }(t)$. $Z_{d}$ satisfies the equation
\begin{equation}
Z_{d}(t)=1-\int\nolimits_{-\infty }^{t}\stackrel{.}{\alpha }(\tau
)\,X_{b}(\tau )d\tau ,  \label{Eq44}
\end{equation}%
which is the formal solution of Eq.(\ref{Eq27}). Therefore the change of the
dark state amplitude $Z_{d}$ is nonlinear in $\stackrel{.}{\alpha }(t)$ and
can be presented in products of $\stackrel{.}{\alpha }$-s. The first
contribution of $\stackrel{.}{\alpha }(t)$\ to $Z_{d}$ can be found as the
square of $\stackrel{.}{\alpha }$, i.e., in the second term of the
expansion. Substituting the corrected $Z_{d}$ into Eqs. (\ref{Eq28}),(\ref%
{Eq29}) (instead of $Z_{d}=1$), one can find the next term in the expansion
of $X_{b}$ and $Y_{c}$. Then the substitution of the thus found $X_{b}$ into
Eq. (\ref{Eq44}) gives the next term of the expansion of $Z_{d}$, etc. In
this paper we consider only the linear corrections to $Y_{b}$ and $X_{c}$.

Our solution, given by the equations (\ref{Eq42})--(\ref{Eq44}), differs
from Crisp's solution in the $B$-time dependence. He considered the case
when $B$ is constant and $\stackrel{.}{\alpha }$ is time dependent. Fig. 4
(a--c) shows the comparison of the numerical solution of the equations (\ref%
{Eq27})--(\ref{Eq29}) with Eqs. (\ref{Eq42})--(\ref{Eq44}). They are\
indistinguishable and shown by the same solid lines.

We find the adiabatic and nonadiabatic components of the analytical solution
by applying two different procedures. The adiabatic components of the
equations (\ref{Eq42})--(\ref{Eq43}) are calculated by integrating them by
parts. For example, the first step of the $X_{c}(t)$ calculation is%
\begin{equation}
X_{c}(t)=\int\nolimits_{-\infty }^{t}\frac{\stackrel{.}{\alpha }(\tau )}{%
B(\tau )}d\left( \cos \left[ \int_{\tau }^{t}B(\tau _{1})d\tau _{1}\right]
\right) =\frac{\stackrel{.}{\alpha }}{B}-\int\nolimits_{-\infty }^{t}\cos %
\left[ \int_{\tau }^{t}B(\tau _{1})d\tau _{1}\right] \left( \frac{\stackrel{.%
}{\alpha }}{B}\right) _{\tau }^{\prime }d\tau .  \label{Eq45}
\end{equation}%
where $\left( \stackrel{.}{\alpha }/B\right) _{\tau }^{\prime }$ is the
derivative with respect to $\tau $. Repeating these steps several times, we
obtain%
\begin{equation}
Y_{b}(t)=\alpha _{2}-\alpha _{4}+\alpha _{6}-\alpha _{8}+...,  \label{Eq46}
\end{equation}%
\begin{equation}
X_{c}(t)=\alpha _{1}-\alpha _{3}+\alpha _{5}-\alpha _{7}+...,  \label{Eq47}
\end{equation}%
where $\alpha _{n}=\stackrel{.}{\alpha }_{n-1}/B$, $\alpha _{0}=\alpha $ and
it is assumed that $\stackrel{.}{\alpha }(t)$ is a bell-shaped function with
first and higher derivatives equal zero at $t=-\infty $. To find a similar
expansion of equation (\ref{Eq44}), we introduce the function $\Omega
(t)=\int\nolimits_{0}^{t}B(\tau )d\tau $. Then the equation for $Z_{d}(t)$
takes the form%
\begin{equation}
Z_{d}(t)=1-A_{c}(t)-A_{s}(t),  \label{Eq48}
\end{equation}%
with%
\begin{equation}
A_{c}(t)=\int\nolimits_{-\infty }^{t}d\tau _{1}\stackrel{.}{\alpha }(\tau
_{1})\cos \Omega (\tau _{1})\int_{-\infty }^{\tau _{1}}d\tau _{2}\stackrel{.}%
{\alpha }(\tau _{2})\cos \Omega (\tau _{2}),  \label{Eq49}
\end{equation}%
\begin{equation}
A_{s}(t)=\int\nolimits_{-\infty }^{t}d\tau _{1}\stackrel{.}{\alpha }(\tau
_{1})\sin \Omega (\tau _{1})\int_{-\infty }^{\tau _{1}}d\tau _{2}\stackrel{.}%
{\alpha }(\tau _{2})\sin \Omega (\tau _{2}).  \label{Eq50}
\end{equation}%
The functions $A_{c}(t)$ and $A_{s}(t)$ are reduced to the single integrals%
\begin{equation}
A_{c}(t)=\frac{1}{2}\left( \int\nolimits_{-\infty }^{t}d\tau _{1}\stackrel{.}%
{\alpha }(\tau _{1})\cos \Omega (\tau _{1})\right) ^{2},  \label{Eq51}
\end{equation}%
\begin{equation}
A_{s}(t)=\frac{1}{2}\left( \int\nolimits_{-\infty }^{t}d\tau _{1}\stackrel{.}%
{\alpha }(\tau _{1})\sin \Omega (\tau _{1})\right) ^{2}.  \label{Eq52}
\end{equation}%
This can be done since they have the structure%
\begin{equation}
A_{c,s}(t)=\int\nolimits_{-\infty }^{t}d\tau \stackrel{.}{f}_{c,s}(\tau
)f_{c,s}(\tau )=\frac{1}{2}f_{c,s}^{2}(t),  \label{Eq53}
\end{equation}%
where $f_{c,s}(\tau )$ is
\begin{equation}
f_{c,s}(t)=\int\nolimits_{-\infty }^{t}d\tau \stackrel{.}{\alpha }(\tau )%
%TCIMACRO{\QATOPD\{ \} {\cos \Omega (\tau )}{\sin \Omega (\tau )}}%
%BeginExpansion
{\cos \Omega (\tau ) \atopwithdelims\{\} \sin \Omega (\tau )}%
%EndExpansion
,  \label{Eq54}
\end{equation}%
and the index $c$ stands for the cosine function and index $s$ for sine.
Integrating these integrals by parts, we obtain%
\begin{equation}
Z_{d}(t)=1-\frac{1}{2}\left[ \left( \alpha _{1}-\alpha _{3}+\alpha
_{5}+...\right) ^{2}+\left( \alpha _{2}-\alpha _{4}+\alpha _{6}+...\right)
^{2}\right] ,  \label{Eq55}
\end{equation}

Equations (\ref{Eq46},\ref{Eq47}) and (\ref{Eq55}) give the probability
amplitudes of the dark, bright and common states. They coincide with those
one obtains if the successive transformations $S_{n}S_{n-1}...S_{1}S$ to the
new set of $d_{n}b_{n}c_{n}$ -states are performed, as was done by
Fleischhauer and coauthors in Ref. \cite{Fleischh99}, i.e., $\left| \Phi
_{n}\right\rangle =S_{n}S_{n-1}...S_{1}S\left| \Phi _{0}\right\rangle $,
where $\left| \Phi _{0}\right\rangle =C_{1}\left| 1\right\rangle
+C_{2}\left| 2\right\rangle +C_{3}\left| 3\right\rangle $ is the initial
state [$C_{1}(-\infty )=1$, $C_{2}(-\infty )=C_{3}(-\infty )=0$]. The
transformation $S$ is defined in Eq. (\ref{Eq17}) and the other
transformations ($S_{i}$) are specified below. These transformations have a
simple meaning. Since the $d-b$ and $b-c$ transitions are excited
simultaneously by the $\stackrel{.}{\alpha }$ and $B$ ``fields'', one can
make a $S_{1}$-transformation to the new set of dark, bright and common
states, $d_{1}$, $b_{1}$ and $c_{1}$, where $d_{1}$ is a particular mixture
of the former $d$ and $c$ states, $b_{1}$ is a state orthogonal to state $%
d_{1}$, and $c_{1}$ coincides with state $b$. The new mixing angle is $%
\alpha _{1}=arc\tan (\stackrel{.}{\alpha }/B)$. Repeating this procedure $n$%
-times, one can get our solution if the conditions $arc\tan (\stackrel{.}{%
\alpha }_{n}/B)\approx \stackrel{.}{\alpha }_{n}/B$ and $\sqrt{\stackrel{.}{%
\alpha }_{n}^{2}+B^{2}}\approx B$ are applied at each step.

Figure 4 (a-c) shows the comparison of the numerical solution (solid lines)
with the expansions given by the equations (\ref{Eq46}), (\ref{Eq47}) and (%
\ref{Eq55}) (dashed lines). The parameters of the pulse train are specified
above and they are the same as in Fig. 3 (a,b). Only the first two terms of
the expansions are taken into account for each plot, which is justified
because $\left| \alpha _{1}\right| \gg \left| \alpha _{2}\right| \gg \left|
\alpha _{3}\right| \gg \left| \alpha _{4}\right| \gg ...$. For the $X_{c}$
and $Y_{b}$ components, the maximum absolute values of the second terms of
the expansions (i.e., the adiabatic terms) are already comparable with the
amplitude of the oscillations (i.e., the nonadiabatic contribution, coming
from the summation of an infinite number of the expansion terms). Of course%
{\it , nonadiabatic oscillations are not described by the main part of the
adiabatic solution presented by a few leading terms of the expansion}.

Concluding this section, we refer to a particular case when $B(t)$ and $%
\stackrel{.}{\alpha }$ have the same time dependence. This is again the case
of matched pulses (see the beginning of Section III), however in the $dbc$
basis. Fleischhauer and coauthors, Ref. \cite{Fleischh99}, classify this
case as second order matched pulses. The solution of the equations (\ref%
{Eq27})--(\ref{Eq29}) is trivial since these equations in terms of a new
variable $\zeta =\int_{-\infty }^{t}F(\tau )d\tau $ can be reduced to a set
of differential equations with constant coefficients, where $%
F(t)=B(t)/B(t_{0})=\stackrel{.}{\alpha }(t)/\stackrel{.}{\alpha }(t_{0})$.
For the first time, this analytically solvable model was considered by
Vitanov and Stenholm in Ref. \cite{VitSten97}. We would classify the case as
nonadiabatic, however in the second order $d_{1}b_{1}c_{1}$-basis, where the
transition takes place. If the generalized pulse area is properly chosen in
this basis, the population transfer between the diabatic states $1$ and $2$
is complete. For these particular pulse areas, there are no nonadiabatic
corrections, which is typical for\ the resonant nonadiabatic transitions.

\section{NONADIABATIC CORRECTIONS}

All adiabatic terms tend to zero at $t\rightarrow +\infty $, which secures
for the three-level atom the adiabatic following of a particular state
coinciding with state $d$ at $t\rightarrow +\infty $. However, as it will be
shown below, if we sum all these infinitely small terms, the result will be
finite. The net value of the small contributions of each adiabatic term is a
nonadiabatic contribution, which specifies the excited probability amplitude
left by the pulse train. To estimate this value, we rewrite the solution,
Eqs. (\ref{Eq42}) and (\ref{Eq43}), as follows%
\begin{equation}
Y_{b}(t)=f_{c}(t)\cos \Omega (t)+f_{s}(t)\sin \Omega (t),  \label{Eq56}
\end{equation}%
\begin{equation}
X_{c}(t)=f_{c}(t)\sin \Omega (t)-f_{s}(t)\cos \Omega (t).  \label{Eq57}
\end{equation}%
When $t\rightarrow +\infty $, the function $f_{c}(t)$ tends to a finite
value whereas $f_{s}(t)$ tends to zero since in the corresponding integrals
[see Eq. (\ref{Eq54})] $\stackrel{.}{\alpha }(\tau )\,$ is an even function
of time and $\Omega (\tau )$ is an odd function. As a result, at the end of
the pulse train $Y_{b}(t)$ and $X_{c}(t)$ oscillate as $\cos \Omega (t)$ and
$\sin \Omega (t)$, respectively, and they have constant amplitudes $%
f_{c}(+\infty )$. The value of $f_{c}(+\infty )$ defines the probability
amplitude left by the pulse sequence in states $b$ and $c$.

The excitation process is adiabatic if the nonadiabatic part $f_{c}(+\infty
) $ is small. If the nonadiabatic part becomes comparable with the main term
$\alpha _{1}(t_{0})$ of the adiabatic expansion, then we can not describe
the excitation process as adiabatic. Figure 5 (a) shows the numerically
found dependences of the $\alpha _{1}(t_{0})$ term (dashed line) and the
nonadiabatic contribution $f_{c}(+\infty )$ (solid line) on the maximum
amplitude $B_{0}$ of the pulses for the pulse train with $Tr=5$. The
adiabatic and nonadiabatic parts of the solution become comparable if $%
B_{0}<15r$.

The semilogarithmic plot of $f_{c}(+\infty )$ versus $B_{0}$, Figure 5 (a),
clearly demonstrates that the nonadiabatic part decreases exponentially with
the increase of $B_{0}$. Laine and Stenholm \cite{Stenholm96} also found an
exponential decrease of the nonadiabatic deviation from the ideal population
transfer with increase of $B_{0}T$\ ($B_{0}$ is the amplitude of each pulse
at the maximum and $T$ is the interpulse distance, fixed in our case by the
relation $Tr=5$, so $B_{0}/r$ is a variable). They followed the calculation
procedure proposed by Davis and Pechukas \cite{Pechukas76} employing the
Dykhne model \cite{Dykhne61-62}. According to Ref. \cite{Pechukas76}, one
has to find the zeros of the function $B(t)$ in a complex plane $t$ and take
the one, $t_{c}$, that is nearest to the real axis. Then the population of
the dark state after the pulse train is%
\begin{equation}
\left| d(+\infty )\right| ^{2}\varpropto \exp \left[ -2%
%TCIMACRO{\func{Im}}%
%BeginExpansion
\mathop{\rm Im}%
%EndExpansion
\int_{t_{0}}^{t_{c}}B(t)dt\right] .  \label{Eq58}
\end{equation}

We propose another calculation procedure of the nonadiabatic deviation. It
will be shown that there are two nonadiabatic contributions, one coming from
the Rabi frequency $B(t)$ and another from the mixing parameter derivative $%
\stackrel{.}{\alpha }(t)$. Only the cooperative contribution of both
determines the net nonadiabatic correction, while the Pechukas-Dykhne
theory, taking into account only the $B(t)$-change, underestimates the
nonadiabatic correction.

To show this, we express $f_{c}(t)$ via the Fourier transform of $\stackrel{.%
}{\alpha }(t)$:%
\begin{equation}
a(\omega )=\int\nolimits_{-\infty }^{+\infty }d\tau \stackrel{.}{\alpha }%
(\tau )\,e^{i\omega \tau },  \label{Eq59}
\end{equation}%
\begin{equation}
f_{c}(t)=\int\nolimits_{-\infty }^{t}d\tau \cos \Omega (\tau )\frac{1}{2\pi }%
\int\nolimits_{-\infty }^{+\infty }d\omega \,a(\omega )\,e^{-i\omega \tau }.
\label{Eq60}
\end{equation}%
Let us consider first the case if the Raman Rabi frequency is constant,
i.e., $B(t)=\beta _{0}=const$. Then $\Omega (\tau )=\beta _{0}\tau $ and%
\begin{equation}
f_{c}(+\infty )=\int_{-\infty }^{+\infty }d\omega \frac{1}{2}\left[ \delta
(\omega +\beta _{0})+\delta (\omega -\beta _{0})\right] \,a(\omega
)=a(\omega )|_{\omega =\beta _{0}},  \label{Eq61}
\end{equation}%
where $\delta (x)$ is the Dirac delta function and $a(\omega )$ is an even
function of $\omega $. The Fourier transform of the mixing parameter
derivative for a secant hyperbolic pulse train can be found, for example in
Ref. \cite{Bateman54}. This function is%
\begin{equation}
a(\omega )=\pi \frac{\sinh \left[ \frac{\omega }{2r}arc\tan \left( \sinh
(rT)\right) \right] }{\sinh \left( \frac{\pi \omega }{2r}\right) }.
\label{Eq62}
\end{equation}%
If we take $\beta _{0}=B(t_{0})=\sqrt{2}B_{0}\sec h\left( rT/2\right) $,
which is the value of the Raman Rabi frequency at time $t=t_{0}$ when $%
\stackrel{.}{\alpha }(t)$ takes its maximum, then for our numerical example
specified above we obtain the amplitude of the nonadiabatic contribution $%
f_{c}(+\infty )=1.263\times 10^{-3}$. This value is four times smaller than
the amplitude of the nonadiabatic oscillations on the right tail of the
functions $X_{c}(t)$ and $Y_{b}(t)$ (which is $\sim 5\times 10^{-3}$), shown
in Figures 4 (a,b), i.e., four times smaller than the nonadiabatic
contribution given by the numerical calculations of the solution of Eqs. (%
\ref{Eq27})--(\ref{Eq29}).

To explain this difference and clarify the origin of the nonadiabatic
contribution, we recall the interaction Hamiltonian in the $dbc$-basis, Eq. (%
\ref{Eq25}) [see also Eqs. (\ref{Eq15}) and (\ref{Eq26})]. This Hamiltonian
resembles the interaction representation Hamiltonian of the three-level
system excited by two resonant ``fields'' with amplitudes $B(t)$ and $%
\stackrel{.}{\alpha }(t)$ (see Fig. 2). Assume, first, that the couplings $%
B(t)$ and $\stackrel{.}{\alpha }(t)$ are absent and the system is in state $%
d $. Then, states $d$, $b$ and $c$ can be considered as having the same
energies in the interaction representation. Switching on the coupling $%
\stackrel{.}{\alpha }(t)=a{}\,\Theta (t)$ [here $\Theta (t)$ is the
Heaviside step function and $a$ is an arbitrary constant] mixes states $d$
and $b$ or in other words induces the transition $d\rightarrow b$. If the $B$%
--field is also present and its amplitude $\beta _{0}$ is constant, i.e., $%
B(t)=\beta _{0}\Theta (t)$, this $B$--field mixes states $b$ and $c$
producing a new couple of states $b^{\prime }$ and $c^{\prime }$, which are
the states $\left| \overline{1}\right\rangle $ and $\left| \overline{3}%
\right\rangle $ [see Eqs. (\ref{Eq8}) and (\ref{Eq10})]. This couple is
split by the energy gap $2\beta _{0}=2B=2\sqrt{B_{1}^{2}+B_{2}^{2}}$ [see
Eq. (\ref{Eq6})]. This is the so called Mollow splitting \cite{Mollow} or
quasi-energy splitting \cite{Zeld73},\cite{Shakhmuratov77}. Assume that
without the $B$--field the $\stackrel{.}{\alpha }$--field is in resonance
with levels $d$ and $b$. The switching on of the $B$--field mixes levels $b$
and $c$, producing an additional splitting. Level $b$ moves on the frequency
$B$ out of resonance with the $\stackrel{.}{\alpha }$--field. If the $%
\stackrel{.}{\alpha }$--field would have a $\delta $--spectrum (in the case
specified above it has only a zero frequency component), then it would not
interact with the atom. However, because of the finite spectral width of the
$\stackrel{.}{\alpha }$--field, its spectrum has a component with frequency $%
B$ on the far tail which is in resonance with the new position of level $b$.
Only this spectral component excites the atom if the $B$--field is on. With
increase of the $B$--field, the component of the $\stackrel{.}{\alpha }$%
--field spectrum, which interacts with the atom, shifts further to the tail
of the spectrum. If the $B$--field amplitude changes in time, several
spectral components of the $\stackrel{.}{\alpha }$--field interact with the
atom since at each instant of time some particular spectral component is in
resonance. The process of the sweeping of the splitting $B(t)$ along the
tail of the $\stackrel{.}{\alpha }$--field spectrum involves a broader band
of the $\stackrel{.}{\alpha }$--spectrum in the interaction. To find the net
atom excitation in this case, we have to calculate the integral $%
f_{c}(+\infty )$ where the change of $B(t)$ is taken into account. This is
done in the next two sections.

Concluding this section, we apply our spectral method to find the
approximate solution of the Rosen-Zener model. The comparison of our
solution with the exact analytical solution helps to find the condition when
our approximation is valid. Referring to the Bloch equations (\ref{Eq30})--(%
\ref{Eq32}) discussed in Section III, we define the splitting of the
two-level system as $\Delta =const$ and the coupling field amplitude as $%
\chi =2V_{0}\sec h(rt)$, where $V_{0}$ is the amplitude of the secant
hyperbolic pulse of width $r$. According to the exact solution (see, for
example, Ref. \cite{Suominen95}), the population of the excited state $e$
after the pulse is%
\begin{equation}
\rho _{ee}(+\infty )=\sin ^{2}\left( \frac{\pi V_{0}}{r}\right) \sec
h^{2}\left( \frac{\pi \Delta }{2r}\right) ,  \label{Eq63}
\end{equation}%
if initially the system is in the ground state $g$, i.e., $\rho
_{gg}(-\infty )=1$.

Employing the correspondence between the Bloch equations for the two-level
system and the Schr\"{o}dinger equation for the three-level system discussed
in Section III, we make the substitution $\stackrel{.}{\alpha }=\chi $ and $%
B=\Delta $ in Eq. (\ref{Eq54}) where $\Omega (t)$ becomes a simple function
of time, i.e., $\Omega (t)=\Delta t$. Then, if $t\rightarrow +\infty $, the
solution is $Y_{b}(t)=f_{c}(+\infty )\cos \Delta t$, $X_{c}(t)=f_{c}(+\infty
)\sin \Delta t$ [see Eqs. (\ref{Eq56})--(\ref{Eq57}) and the discussion
following these equations]. $f_{s}(+\infty )=0$ since $\Omega (t)$ is the
odd function. $f_{c}(+\infty )$ is simply the Fourier transform $\chi
(\Delta )$ [defined as a nondimensional value according to Eqs. (\ref{Eq59})
and (\ref{Eq61})] of the function $\chi (t)$ at the frequency $B=\Delta $,
i.e., $f_{c}(+\infty )=\chi (\Delta )$, [see Eqs. (\ref{Eq61}) and (\ref%
{Eq59})], or explicitly%
\begin{equation}
f_{c}(+\infty )=2V_{0}\int\nolimits_{-\infty }^{+\infty }dt\sec h(rt)\,\cos
\Delta t.  \label{Eq64}
\end{equation}%
This integral also can be found in Ref. \cite{Bateman54}: $f_{c}(+\infty
)=2\pi (V_{0}/r)\sec h(\pi \Delta /2r)$.

The population of the excited state $e$ of the two-level system can be
expressed as follows $\rho _{ee}=(1-w)/2$. Since there is a reciprocity
between the Bloch-vector components of the two-level system and the state
probability amplitudes of the three-level system ($w=Z_{d}$, $v=Y_{b}$, $%
u=X_{c}$), the excited state population is $\rho _{ee}=(1-Z_{d})/2$. The
probability amplitudes are normalized, $X_{c}^{2}+Y_{b}^{2}+Z_{d}^{2}=1$,
and hence $Z_{d}=\sqrt{1-X_{c}^{2}-Y_{b}^{2}}=\sqrt{1-f_{c}^{2}(+\infty )}$.
Assuming that $f_{c}(+\infty )\ll 1$, we obtain%
\begin{equation}
\rho _{ee}(+\infty )\simeq \left( \frac{\pi V_{0}}{r}\right) ^{2}\sec
h^{2}\left( \frac{\pi \Delta }{2r}\right) .  \label{Eq65}
\end{equation}%
Thus, our approximate result, Eq. (\ref{Eq65}), coincides with the exact
solution, Eq. (\ref{Eq63}), if the population change is small and the sine
function can be represented by the first term of its expansion in a power
series. These conditions specify the validity of the adiabatic
approximation. What is remarkable here is that the dependence of the excited
state population on the two-level splitting $\Delta $ in the approximate
solution coincides with that in the exact solution. This shows that our
spectral approach in the description of the nonadiabatic corrections is
valid.

If $\Delta \rightarrow 0$, the nonadiabatic transition between states $g$
and $e$ takes place according to the straightforward solution of the
Bloch-equation for the two-level system: $\rho _{ee}(t)=\sin ^{2}[\theta
(t)/2]$, where $\theta (t)=\int_{-\infty }^{t}\chi (\tau )d\tau $. At $%
t\rightarrow +\infty $ we have $\rho _{ee}(+\infty )=\sin ^{2}(\pi V_{0}/r)$%
, which reproduces the exact Rosen-Zener solution (\ref{Eq63}) if $\Delta
\rightarrow 0$.

\section{NONADIABATIC TRANSITION AS A QUANTUM JUMP: BASIC ARGUMENTS}

If the Raman Rabi frequency changes during the development of the mixing
parameter $\stackrel{.}{\alpha }(t)$, those Rabi frequencies sweeping the
frequency bandwidth of $\stackrel{.}{\alpha }(t)$ contribute to the
nonadiabatic corrections. To take this process into account, we have to
convolute the spectral content of both the mixing parameter and the Raman
Rabi frequency. In general, this calculation is nontrivial. For the case of
secant hyperbolic pulses [see the time development of the pulses, the Raman
Rabi frequency and the mixing parameter in Fig. 3 (a,b)], it is possible to
simplify the problem. Since $B(t)$ has a minimum value of $\beta _{0}$ at $%
t_{0}$ where $\stackrel{.}{\alpha }(t)$ has its maximum, only the spectral
components of $a(\omega )$ with $\left| \omega \right| \geqslant \beta _{0}$
contribute. Because time and frequency domains are inversely proportional,
the main part of the nonadiabatic contribution appears in a short time
interval around $t_{0}$, corresponding to the far tail of the spectral
distribution $a(\omega )$. This means that the time scale of the
nonadiabatic change is shorter than $1/\beta _{0}$, while, according to the
condition imposed on the parameters, the time variation of $\stackrel{.}{%
\alpha }(t)$ is much longer than the Rabi oscillation period $\sim 1/\beta
_{0}$. Thus, a nonadiabatic transition takes place almost jumpwise compared
to the time scale of the variation of $\stackrel{.}{\alpha }(t)$. Due to
this circumstance, we can simplify the calculation of the integral $%
f_{c}(+\infty )$ by expanding $B(t)$ in a power series of $t$ near $t_{0}$
and retaining only the first two terms of the expansion%
\begin{equation}
B(t)\approx \beta _{0}\left[ 1+g(t-t_{0})^{2}\right] ,  \label{Eq66}
\end{equation}%
where $g=r^{2}\left[ 3\tanh ^{2}(rT/2)-1\right] /2$. We verified this
approximation by comparing numerically two integrals $f_{c}(+\infty )$ and $%
f_{cA}(+\infty )$ calculated with $B(t)$ and its approximated value, Eq. (%
\ref{Eq66}), respectively. The comparison is shown in Fig. 5 b, where $%
f_{cA}(\infty )$ (dashed line) is the approximation. The dependences of both
integrals on the amplitude $B_{0}$ are indistinguishable.

The approximation of $B(t)$ by a parabolic function helps to express $%
f_{c}(+\infty )$ via the Airy integral. Further, to simplify the notations
we set $t_{0}=0$. Then the phase $\Omega (t)$ is%
\begin{equation}
\Omega (t)=\beta _{0}\left( t+\frac{g}{3}t^{3}\right) ,  \label{Eq67}
\end{equation}%
and $f_{cA}(+\infty )$ is expressed as%
\begin{equation}
f_{cA}(+\infty )=\frac{1}{2\pi }\int_{-\infty }^{+\infty }d\tau \cos \left[
\beta _{0}\left( \tau +\frac{g}{3}\tau ^{3}\right) \right] \int_{-\infty
}^{+\infty }d\omega \,a(\omega )e^{-i\omega \tau }.  \label{Eq68}
\end{equation}%
Evaluating the time integral in Eq. (\ref{Eq68}), we obtain%
\begin{equation}
f_{cA}(+\infty )=\gamma \int_{-\infty }^{+\infty }d\omega \,a(\omega )Ai
\left[ \gamma (\beta _{0}+\omega )\right] ,  \label{Eq69}
\end{equation}%
where $Ai(x)$ is Airy integral and $\gamma =1/\sqrt[3]{g\beta _{0}}$. We
derived this equation assuming that $a(\omega )$ is an even function. The
Airy integral has a different dependence for positive and negative arguments
(see, for example Ref. \cite{AbramSteg65}). If $\beta _{0}+\omega \geq 0$,
we have%
\begin{equation}
\gamma Ai\left[ \gamma (\beta _{0}+\omega )\right] =\frac{1}{3}\sqrt{\frac{%
\beta _{0}+\omega }{3g\beta _{0}}}\left\{ I_{-\frac{1}{3}}\left[ \frac{2}{3}%
\sqrt{\frac{(\beta _{0}+\omega )^{3}}{g\beta _{0}}}\right] -I_{\frac{1}{3}}%
\left[ \frac{2}{3}\sqrt{\frac{(\beta _{0}+\omega )^{3}}{g\beta _{0}}}\right]
\right\} ,  \label{Eq70}
\end{equation}%
where $I_{\pm \frac{1}{3}}(x)$ is the modified Bessel function of order $\pm
\frac{1}{3}$. If the argument is negative, $\beta _{0}+\omega <0$, then%
\begin{equation}
\gamma Ai\left[ \gamma (\beta _{0}+\omega )\right] =\frac{1}{3}\sqrt{\frac{%
\left| \beta _{0}+\omega \right| }{3g\beta _{0}}}\left\{ J_{-\frac{1}{3}%
}\left( \frac{2}{3}\sqrt{\frac{\left| \beta _{0}+\omega \right| ^{3}}{g\beta
_{0}}}\right) +J_{\frac{1}{3}}\left( \frac{2}{3}\sqrt{\frac{\left| \beta
_{0}+\omega \right| ^{3}}{g\beta _{0}}}\right) \right\} ,  \label{Eq71}
\end{equation}%
where $J_{\pm \frac{1}{3}}(x)$ is the Bessel function of order $\pm \frac{1}{%
3}$. For large arguments both functions have a simple asymptotic behavior%
\begin{equation}
\gamma Ai\left[ \gamma (\beta _{0}+\omega )\right] \approx \frac{\exp \left[
-\frac{2}{3}\sqrt{\frac{(\beta _{0}+\omega )^{3}}{g\beta _{0}}}\right] }{2%
\sqrt{\pi }\left[ g\beta _{0}(\beta _{0}+\omega )\right] ^{\frac{1}{4}}},
\label{Eq72}
\end{equation}%
if $\beta _{0}+\omega $ is positive, and%
\begin{equation}
\gamma Ai\left[ \gamma (\beta _{0}+\omega )\right] \approx \frac{\cos \left(
\frac{2}{3}\sqrt{\frac{\left| \beta _{0}+\omega \right| ^{3}}{g\beta _{0}}}-%
\frac{\pi }{4}\right) }{\sqrt{\pi }\left( g\beta _{0}\left| \beta
_{0}+\omega \right| \right) ^{\frac{1}{4}}},  \label{Eq73}
\end{equation}%
if $\beta _{0}+\omega $ is negative. If the argument is zero, $\beta
_{0}+\omega =0$, this function is%
\begin{equation}
\gamma Ai\left[ \gamma (\beta _{0}+\omega )\right] |_{\omega =-\beta _{0}}=%
\frac{\Gamma (\frac{1}{3})}{2\pi \sqrt[3]{\sqrt{3}g\beta _{0}}},
\label{Eq74}
\end{equation}%
where $\Gamma (\frac{1}{3})\approx 2.679$ is the gamma function.

\section{RABI CHIRPING ONLY}

If the $\stackrel{.}{\alpha }$-field does not depend on time and has a value
$\stackrel{.}{\alpha }(t)=\stackrel{.}{\alpha }(t_{0})$, then $a(\omega
)=2\pi \stackrel{.}{\alpha }(t_{0})\delta (\omega )/r$ and the main
contribution to the nonadiabatic part is given by the Airy integral%
\begin{equation}
f_{cA}(+\infty )=2\pi \gamma Ai(\gamma \beta _{0})\stackrel{.}{\alpha }%
(t_{0})/r,  \label{Eq75}
\end{equation}%
which has the explicit form
\begin{equation}
f_{cA}(+\infty )=\tanh \left( \frac{rT}{2}\right) \sqrt{\frac{\pi }{\beta
_{0}\sqrt{g}}}\exp \left( -\frac{2}{3}\frac{\beta _{0}}{\sqrt{g}}\right) ,
\label{Eq76}
\end{equation}%
expressed via approximation (\ref{Eq72}). In this section we show how the
approximation, specified above, is related to the Dykhne-Pechukas model \cite%
{Dykhne61-62},\cite{Pechukas76}.

For large $\beta _{0}$, the Airy integral can be calculated by the {\it %
saddle point methods} as shown, for example, in Ref. \cite{Walker65}. First,
the stationary or saddle point $t_{s}$ is found where the phase $\Omega (t)$
becomes stationary: $\stackrel{.}{\Omega }(t_{s})=0$. In this point, the
Raman Rabi frequency becomes zero since $\stackrel{.}{\Omega }(t)=B(t)$. For
the case of the positive argument of the Airy integral, the saddle point is
in the complex plane and $t_{s}$ has only an imaginary component [$%
%TCIMACRO{\func{Re}}%
%BeginExpansion
\mathop{\rm Re}%
%EndExpansion
(t_{s})=0$], i.e.,%
\begin{equation}
t_{s}=-\frac{i}{\sqrt{g}}=-\frac{i}{r}\sqrt{\frac{2}{3\tanh ^{2}(rT/2)-1}}.
\label{Eq77}
\end{equation}%
In fact, there are two saddle points: $t_{s}=\pm i/\sqrt{g}$. The negative
sign is chosen to avoid exponentially increasing numbers. Then the method of
stationary phase (one of the saddle point{\it \ }methods) is applied. This
method is applicable to integrals of the form%
\begin{equation}
I(\beta _{0})=\int_{C}e^{i\beta _{0}f(z)}dz,
\end{equation}%
where $\beta _{0}$ is large and $C$ is a path in the complex plane such that
the ends of the path do not contribute significantly to the integral. The
idea of the method is to deform the contour $C$ so that the region of most
of the contribution to $I(\beta _{0})$ is compressed into as short a space
as possible. This compression occurs at the saddle point.

Since the main contribution to this integral comes from the vicinity of the
saddle point, one may conclude that in the adiabatic limit the atom abruptly
makes a nonadiabatic transition between adiabatic states. This is very much
like crossing a Stokes line for the asymptotic series, and resembles the
turning point problem in the WKB theory (see, for example, Refs. \cite%
{Suominen95}, \cite{Walker65}). In the WKB method, the one-dimensional Schr%
\"{o}dinger equation is considered for a particle with mass $m$ in a
potential $V(x)$:%
\begin{equation}
\frac{d^{2}\Psi }{dx^{2}}=-\frac{2m}{\hbar ^{2}}[E-V(x)]\Psi .  \label{Eq78}
\end{equation}%
The turning point is the one where the energy of the particle coincides with
the potential: $E-V(x)=0$. If $2m[E-V(x)]/\hbar ^{2}\propto x$, the solution
of the equation (\ref{Eq78}) can be expressed via the Airy integral \cite%
{Walker65}. In this approach the WKB connection formula, relating the
exponentially small solution on one side of a turning point to an
oscillatory solution on the other side, is derived quite naturally.

The Pechukas-Dykhne \cite{Pechukas76} recipe of the calculation of the
nonadiabatic contribution is similar to the method described above since the
definitions of the saddle point $t_{s}$ and the crossing point $t_{c}$,
where $B(t_{c})$ is zero, are the same. However, in our consideration we
substituted the expression for the Raman Rabi frequency%
\begin{equation}
B(t)=2B_{0}\frac{\sqrt{1+\cosh (rT)\cosh (2rt)}}{\cosh (rT)+\cosh (2rt)},
\label{Eq79}
\end{equation}%
by the expansion near time $t_{0}$, Eq. (\ref{Eq66}). Therefore, we have
only two saddle or crossing points, (\ref{Eq78}), whereas equation (\ref%
{Eq79}) has an infinite number of crossing points in the complex plane.
According to Pechukas, one has to take the crossing point that is nearest to
the real axis. For the secant hyperbolic pulse train, this point, as shown
by Stenholm \cite{Stenholm96}, has only an imaginary part $%
%TCIMACRO{\func{Im}}%
%BeginExpansion
\mathop{\rm Im}%
%EndExpansion
(t_{c})$, i.e.,%
\begin{equation}
t_{c}=\frac{i}{r}\tan ^{-1}\left[ \coth \left( \frac{rT}{2}\right) \right] .
\label{Eq80}
\end{equation}%
Reformulating the Dykhne approach \cite{Dykhne61-62}, we conclude: if there
are no spectral components of $a(\omega )$ matching the frequency gap
between the quasi-energy levels split by the $B$-field, the $\stackrel{.}{%
\alpha }$-field comes to resonance with the $d-c$ transition at the
imaginary time $t_{c}$ when $B$ is zero.

This approach disregards the spectral content of $\stackrel{.}{\alpha }(t)$.
If we take $\stackrel{.}{\alpha }(t)=\stackrel{.}{\alpha }(t_{0})$
throughout the excitation, this method gives also an underestimated value of
the nonadiabatic contribution. For our numerical example, equation (\ref%
{Eq76}) gives $f_{cA}(+\infty )=7.131\times 10^{-4}$, which is $7$ times
smaller than the value of the nonadiabatic contribution given by the
numerical calculation of the Schr\"{o}dinger equation (\ref{Eq27})--(\ref%
{Eq29}). Figure 6 (a) shows the comparison of the numerically calculated
nonadiabatic component $f_{c}(+\infty )$ (solid line), Eq. (\ref{Eq54}),
with that calculated for the case if the time dependence of the $\stackrel{.}%
{\alpha }(t)$-field is neglected (dashed line). The latter is given by Eq.(%
\ref{Eq76}). The plots obviously demonstrate that this approximation
underestimates the nonadiabatic contribution.

Concluding this Section, we demonstrate the application of our method to the
Landau-Zener model. In this model the parameters of the corresponding
two-level system are $\Delta =2\lambda t$ and $\chi =2V_{0}$. Here, the
level crossing takes place at $t=0$ and the nonadiabatic correcton appears
in the vicinity of the crossing point. After the substitution specified in
Section III and repeating the same procedure as described at the end of
Section IV, we find that%
\begin{equation}
f_{c}(+\infty )=2V_{0}\int_{-\infty }^{+\infty }dt\cos (\lambda t^{2}),
\label{Eq81}
\end{equation}%
\begin{equation}
f_{s}(+\infty )=2V_{0}\int_{-\infty }^{+\infty }dt\sin (\lambda t^{2}).
\label{Eq82}
\end{equation}%
The value $f_{s}(+\infty )$ is not zero for this model since $\Omega
(t)=\lambda t^{2}$ is even function of time. The integrals in Eqs. (\ref%
{Eq81})-(\ref{Eq82}) are expressed via Fresnel integral asymptotes (see, for
example, \cite{AbramSteg65}): $f_{c}(+\infty )=f_{s}(+\infty )=V_{0}\sqrt{%
2\pi /\lambda }$. Recall that $Z_{d}(+\infty )=\sqrt{1-f_{c}^{2}(+\infty
)-f_{s}^{2}(+\infty )}$. Then in the adiabatic limit, $V_{0}^{2}\ll \lambda
/2\pi $, we obtain%
\begin{equation}
\rho _{ee}(+\infty )=\frac{1-Z_{d}(+\infty )}{2}\simeq \pi \frac{V_{0}^{2}}{%
\lambda }.  \label{Eq83}
\end{equation}%
This expression coincides with the leading term of the exact solution, Ref. %
\cite{Suominen95}, expansion, that is%
\begin{equation}
\rho _{ee}(+\infty )=1-\exp \left( -\pi \frac{V_{0}^{2}}{\lambda }\right)
\simeq \pi \frac{V_{0}^{2}}{\lambda }.  \label{Eq84}
\end{equation}

We have to emphasize that our solution is valid only if the atom
adiabatically follows the ground state $g$. Therefore, the nonadiabatic
correction (\ref{Eq83}), specifying the population probability of the
excited state $e$, coincides with the Landau-Zener solution, Eq. (\ref{Eq84}%
), if $V_{0}^{2}\ll \lambda /2\pi $, i.e., when the population of the $e$
and $g$ states does not change appreciably. In this case our three-level
system $dbc$, equivalent to the corresponding two-level system $eg$, also
follows adiabatically state $d$. The opposite case is considered in the
Appendix. There the new adiabatic states $d_{1}b_{1}c_{1}$ are introduced
whose population also does not change appreciably. This transformation is
necessary if $V_{0}^{2}\gg \lambda /2\pi $, because the reciprocal
three-level system $dbc$ does not follow state $d$, making a transition from
state $d$ to state $c$ and back at the level crossing.

The necessity of the transformation can also be formulated in a different
way. To apply the adiabatic solution, we need a certain relation between the
interaction parameters $B$ and $\stackrel{.}{\alpha }$. Then the expansion
in $\alpha $'s, where each next term of the expansion is smaller than the
previous one [see Eqs. (\ref{Eq42})-(\ref{Eq55})], is applicable. If $%
V_{0}^{2}\ll \lambda /2\pi $, this expansion can be made in the $dbc$-basis.
In the opposite case, this expansion is not applicable. Therefore, we have
to find new $d_{1}b_{1}c_{1}$-basis where the new interaction parameters $%
\widetilde{B}$ and $\stackrel{.}{\alpha }_{1}$ are related such that the
expansion in $\alpha $'s converges.

\section{COOPERATIVE CONTRIBUTION OF RABI CHIRPING AND THE TIME DEPENDENT
COUPLING, THE APPLICATION TO STIRAP BY SECANT HYPERBOLIC PULSES}

In previous sections we showed that the contribution to the nonadiabatic
corrections of the Rabi frequency chirping or the time dependence of the $%
\stackrel{.}{\alpha }(t)$-field, being taken into account separately, was
not enough to have the right nonadiabatic correction for STIRAP induced by
secant hyperbolic pulses. Only both processes taken together, i.e., the Rabi
frequency $B(t)$ chirping and the participation of all spectral components $%
a(\omega )$ of the $\stackrel{.}{\alpha }(t)$-field engaged by this chirping
to excite the atom, give the right value of the nonadiabatic component. This
value is given by Eq. (\ref{Eq69}). To calculate analytically the
convolution integral of the two spectra, $a(\omega )$ and the Airy integral
describing the process of the Rabi frequency sweeping, we make two
approximations. First, we take the approximation of the Airy integral, given
by Eq. (\ref{Eq72}) in the frequency domain $\omega \geq -\beta _{0}$. This
gives a small overestimation of the integrand near $-\beta _{0}$. For this
reason, we start the integration from this value, not from $-\infty $. The
oscillating part of the Airy integral for $\omega <-\beta _{0}$ gives a much
smaller contribution than the main part, which is located between $-\beta
_{0}$ and zero, $(-\beta _{0},0)$. So, neglecting the part $(-\infty ,-\beta
_{0})$, we compensate the overestimation near $-\beta _{0}$. Second, we
approximate the spectrum $a(\omega )$ in the domain $(-\beta _{0},0)$ by%
\begin{equation}
a(\omega )=\pi \exp \left( -R\left| \omega \right| \right) ,  \label{Eg85}
\end{equation}%
where $R=\left( \pi -arc\tan [\sinh (rT)]\right) /2r\approx \pi /4r$. This
approximation also gives a slight overestimation of the integrand near $%
\omega \sim 0$. To compensate this, we stop the integration at $\omega =0$.
The integrand has a maximum between $\omega =-\beta _{0}$ and $\omega =0$.
To calculate the contribution of this part, we use the modified method of
the saddle point method. Usually, in the method of the saddle point, the
integration near the point is extended to $\pm \infty $. In our case, to
avoid the overestimation of the integrand we limit the integration by finite
boundaries. The calculation of these boundaries for the deformed integration
contour $C$ is simple because the deformed $C$ stays on the real axis. The
result of the integration is%
\begin{equation}
f_{cA}(+\infty )=\pi Q\exp \left[ -\beta _{0}R\left( 1-\frac{1}{3}%
gR^{2}\right) \right] ,  \label{Eq86}
\end{equation}%
where $Q$ is a correction factor, which takes into account the finite
integration boundaries to avoid the overestimation of the integral.
Explicitly, we have%
\begin{equation}
Q=\frac{1}{2}\left[
%TCIMACRO{\func{erf}}%
%BeginExpansion
\mathop{\rm erf}%
%EndExpansion
(h_{\max })+%
%TCIMACRO{\func{erf}}%
%BeginExpansion
\mathop{\rm erf}%
%EndExpansion
(h_{\min })\right] ,  \label{Eq87}
\end{equation}%
\begin{equation}
h_{\max }=\sqrt{\beta _{0}\left[ \frac{2}{3\sqrt{g}}-\left( \frac{2}{3}%
\right) ^{\frac{4}{3}}R+\frac{1}{3}R^{3}g\right] },  \label{Eq88}
\end{equation}%
\bigskip
\begin{equation}
h_{\min }=\sqrt{\frac{\beta _{0}gR^{3}}{3}},  \label{Eq89}
\end{equation}%
and $%
%TCIMACRO{\func{erf}}%
%BeginExpansion
\mathop{\rm erf}%
%EndExpansion
(x)$ is the error function. Figure 6 (b) shows the comparison of the true
nonadiabatic contribution $f_{c}(+\infty )$ (solid line) with our
approximate calculation of $f_{cA}(+\infty )$, given by Eq. (\ref{Eq86})
(dashed line). Figure 6 (c) compares the same dependencies if the correction
factor is $Q=1$ (dash dotted line).

For our numerical example specified above, the approximated value of the
nonadiabatic contribution is $f_{cA}(+\infty )=5.822\times 10^{-3}$. As was
discussed above, this contribution appears in a very short time interval
around $t_{0}$. Therefore, we can approximate the solution of the Schr\"{o}%
dinger equations (\ref{Eq27})--(\ref{Eq29}) by%
\begin{equation}
X_{bA}(t)\approx \alpha _{2}-\alpha _{4}+\Theta (t-t_{0})\,f_{cA}(+\infty
)\cos \Omega (t),  \label{Eq90}
\end{equation}%
\begin{equation}
X_{cA}(t)\approx \alpha _{1}-\alpha _{3}+\Theta (t-t_{0})\,f_{cA}(+\infty
)\sin \Omega (t).  \label{Eq91}
\end{equation}%
where the index $A$ stands for the approximation, $\Theta (t-t_{0})$ is the
Heaviside step function, $\Omega (t)=\int_{0}^{t}B(\tau )d\tau $ and $B$ has
its exact value [not the approximation (\ref{Eq66})]. Figures 7 (a,b) show
the comparison of the numerically found solutions of the Schr\"{o}dinger
equation (solid lines) with the approximation (\ref{Eq90})--(\ref{Eq91})
(dashed lines). The fit of the solutions is striking. This means that the
nonadiabatic contribution really appears in a quite short time range around $%
t_{0}$.

\section{CONCLUSION}

The introduction of the basis of the dark and bright states facilitates the
understanding of the physical processes in the three-level atom excited by a
bichromatic field. The dynamic evolution of the atom between the dark,
bright and common states is as simple as the dynamics of the two-level atom
excited by one field with Rabi frequency $B$. This is because the evolution
of the three-level atom can be effectively reduced to the evolution between
two states, i.e., the bright and common states. The reduction of the
three-level model to the two-level one allows also the application of the
Bloch-vector model and Bloch equations for the treatment of the three-level
atom excitation by the bichromatic field.

This quite simple algebra is applicable for the case of matched pulses. If
the pulses do not match in shape and have different time dependencies, one
can also reduce the consideration to the Bloch-vector model since there is a
similarity between the Schr\"{o}dinger equations for the probability
amplitudes of the dark, bright and common states and the Bloch equation for
an effective two-level system. The effective detuning of the two-level
system from resonance is $B(t)$ and the Rabi frequency is $\stackrel{.}{%
\alpha }(t)$, which is the derivative of the mixing angle in the dark state
development in states of the reciprocal three-level system. This similarity
allows a simple interpretation of the physical processes in the three-level
system in case of adiabatic following of the dark state. In terms of the
Bloch-vector model, the adiabatic following criterion is formulated
resulting in three conditions. First, the initial state must coincide with
the Bloch-vector aligned along an effective field at $t=-\infty $. This
field is determined by the value $\sqrt{B^{2}(t)+\stackrel{.}{\alpha }^{2}(t)%
}$. Second, the effective field makes a small angle with its initial
position when the reciprocal parameter $\stackrel{.}{\alpha }(t)$
responsible for the atom-field interaction takes its maximum. Third, the
time scale of the change of the ''pulse'', determined by the scale of the
variation $\stackrel{.}{\alpha }(t)$, is much longer that the period of
oscillations induced by the effective field $B(t)$. If all three conditions
are met, the dynamic evolution of the three-level atom excited by a sequence
of pulses is described by simple algebra. The adiabatic population transfer
has a nice interpretation in terms of dark, bright and common states.

We developed an approximation describing the adiabatic interaction of the
three-level atom with two resonant pulses. The method of estimation of the
nonadiabatic correction is presented. It is applied to the case of two
secant hyperbolic pulses. The adiabatic part of the solution describes the
excitation and de-excitation processes of the three-level atom. The time
dependence of the adiabatic part is smooth and follows the derivatives of
the mixing angle $\alpha (t)$. Both parts, excitation and de-excitation, are
symmetric in time with respect to $t=0$ where $\stackrel{.}{\alpha }(t)$
takes its maximum and they exactly compensate each other. In this respect
the adiabatic following resembles the soliton-like interaction with the
field. The nonadiabatic part appears in a short time interval in the
vicinity of the maximum of the mixing parameter $\stackrel{.}{\alpha }(t)$.
It contains the information about the excitation left in the atom by the
pulses because of the imperfect following of the dark state. Since this
excitation lasts only a short time, we consider the nonadiabatic process as
a transition between the ground and excited states, which takes place like a
jump. The origin of the transition has a simple interpretation. The
parameter $B$ defines the coupling strength of the bright ($b$) and common ($%
c$) states of the three-level system in the $dbc$--basis. The parameter $%
\stackrel{.}{\alpha }(t)$ defines the coupling strength of the bright and
dark ($d$) states. It is assumed that, initially, the system is in the dark
state and any time we have $B\gg \stackrel{.}{\alpha }(t)$. The $B$%
--coupling moves the bright state from the resonance with the $\stackrel{.}{%
\alpha }(t)$--coupling to the value determined by the frequency $B$. If the
spectral content of the $\stackrel{.}{\alpha }(t)$--coupling has a component
with frequency $B$, the nonadiabatic transition takes place. If $B$ changes
in time, several spectral components of the $\stackrel{.}{\alpha }(t)$%
--coupling contribute to the transition. Our results are compared with those
of Laine-Stenholm \cite{Stenholm96} and Fleischhauer et al. \cite{Fleischh99}%
.

We applied also our method to the Rosen-Zener and Landau-Zener models. The
comparison of our approximate solution with the exact solutions available
for these models is quite encouraging. We obtained exponentially small
nonadiabatic corrections for the case of a symmetric time dependence of the
excitation parameters $B(t)$ and $\stackrel{.}{\alpha }(t)$ with respect to $%
t=0$ (our model, the Rosen-Zener model and the Landau-Zener model with $%
V_{0}^{2}\gg \lambda /2\pi )$, and a small power dependent nonadiabatic
correction for the case of an asymmetric time dependence of $B(t)$
(Landau-Zener model with $V_{0}^{2}\ll \lambda /2\pi $). We considered also
the excitation of the three-level system by asymmetric pulses and found a
power dependent nonadiabatic correction. The results on asymmetric pulses,
exciting the three-level atom, will be presented in another paper.

The simplified algebra developed in this paper could be useful for the
description of the atom state manipulation by coherent fields.

\section{ACKNOWLEDGMENTS}

This work was supported by the Fonds voor Wetenschappelijk Onderzoek
Vlaanderen, the UIAP program of the Belgian government, ISTC (2121) and the
Russian Foundation for Basic Research.

\section{APPENDIX}

In this section we compare our approximate solution with an exact solution
of the Landau-Zener model if the parameters of the corresponding two-level
system $\Delta =2\lambda t$ and $\chi =2V_{0}$, satisfy the condition $%
V_{0}^{2}\gg \lambda /2\pi $. This condition is opposite to the one
considered at the end of Section VII. There, since the coupling $V_{0}$ is
small compared to the sweeping rate $\lambda $ of the two-level system
splitting $\Delta $ ($V_{0}^{2}\ll \lambda /2\pi $), the system
adiabatically follows the ground state $g$ and the net population of the
excited state $e$ is small for $t\rightarrow +\infty $. If $V_{0}^{2}\gg
\lambda /2\pi $, states $g$ and $e$ are not an adiabatic basis. To apply the
adiabatic approximation in this case, we have to choose a new basis that is
adiabatic. This can be done if we use again the similarity between the
two-level and three-level systems discussed in Section II. We consider the
three-level system $dbc$ described by the Hamiltonian (\ref{Eq25}) with the
parameters $B=\Delta =2\lambda t$ and $\stackrel{.}{\alpha }=\chi =2V_{0}$.
Then we make a $S_{1}$--transformation to the new basis $d_{1}b_{1}c_{1}$

\begin{equation}
\left| d_{1}\right\rangle =\sin \alpha _{1}\left| d\right\rangle +i\cos
\alpha _{1}\left| c\right\rangle ,  \label{Eq92}
\end{equation}%
\begin{equation}
\left| b_{1}\right\rangle =i\cos \alpha _{1}\left| d\right\rangle +\sin
\alpha _{1}\left| c\right\rangle ,  \label{Eq93}
\end{equation}%
\begin{equation}
\left| c_{1}\right\rangle =\left| b\right\rangle ,  \label{Eq94}
\end{equation}%
where $\alpha _{1}=arc\tan \left( \Delta /\chi \right) $ and the new Rabi
frequency is $\widetilde{B}=\sqrt{\Delta ^{2}+\chi ^{2}}$. A new mixing
angle $\alpha _{1}$ takes the values $\pm \pi /2$ at $t=\pm \infty $. Thus,
it varies from $-\pi /2$ to $+\pi /2$ with time and, hence, state $d_{1}$
coincides with $-\left| d\right\rangle $ and $+\left| d\right\rangle $
before and after the crossing of levels $g$ and $e$, which takes place at $%
t=0$. If the system adiabatically follows the changing state $d_{1}$, no
transition to the $b_{1}$ state takes place and, finally, the system is left
in the initial state $d$ but with the $\pi $-shift of its phase. This phase
shift originates from the transition $d\rightarrow c\rightarrow d$ in the $%
dbc$--basis. One can find that, according to Eq.(\ref{Eq92}), the system is
in state $d$ at $t=\pm \infty $ when $\alpha _{1}=\pm \pi /2$, and in state $%
c$ at $t=0$ when $\alpha _{1}=0$ if the system follows state $d_{1}$.

We know from the exact solution of the Landau-Zener model that $\rho
_{gg}(+\infty )=\exp (-\pi V_{0}^{2}/\lambda )$ and $\rho _{ee}(+\infty
)=1-\exp (-\pi V_{0}^{2}/\lambda )$ if $\rho _{gg}(-\infty )=1$ [see Eq. (%
\ref{Eq84})]. Thus, if $\pi V_{0}^{2}\gg \lambda $, the population of the
ground state $g$ of the corresponding two-level system is exponentially
small at $t\rightarrow +\infty $. With the help of our method for the
estimation of the nonadiabatic correction, we can find the probability
amplitudes of states $b_{1}$ and $c_{1}$ left by the interaction $\chi $
(coupling states $d$ and $b$) after the level crossing of the equivalent
two-level system. These probability amplitudes allow to estimate the extent
of the depopulation of state $d_{1}$. If the system adiabatically follows
the $d_{1}$ state, the state-vector changes from $-\left| d\right\rangle $
to $+\left| d\right\rangle $. Since the development coefficient $C_{d}$ of
the state vector in the $dbc$--basis corresponds to the $Z$--component of
the Bloch-vector of the equivalent two-level system, the change of the $%
C_{d} $ sign means the flip of the $Z$--component. This flip means a
population inversion of the two-level system, i.e., if before level crossing
(at $t=-\infty $) the system is in the ground state $g$, after the level
crossing (at $t=+\infty $) it is in the excited state $e$. We can estimate
the amount of population left in the ground state, which is the nonadiabatic
correction.

To find the nonadiabatic correction, we apply the $S_{1}$--transformation%
\begin{equation}
S_{1}=\left[
\begin{array}{ccc}
\sin \alpha _{1} & \ 0 & -i\cos \alpha _{1} \\
-i\cos \alpha _{1} & \ 0 & \;\;\sin \alpha _{1} \\
0 & \;0 & \;\;0%
\end{array}%
\right]  \label{Eq95}
\end{equation}%
to the Hamiltonian (\ref{Eq25}) with the parameters $B=2\lambda t$ and $%
\stackrel{.}{\alpha }=2V_{0}$. The result is%
\begin{equation}
\overline{{\cal H}}_{d_{1}b_{1}c_{1}}={\cal H}_{d_{1}b_{1}c_{1}}+i\stackrel{.%
}{S}_{1}S_{1}^{-1},  \label{Eq96}
\end{equation}%
where%
\begin{equation}
{\cal H}_{d_{1}b_{1}c_{1}}={\cal -}\sqrt{\Delta ^{2}+\chi ^{2}}\left(
\widehat{P}_{b_{1}c_{1}}+\widehat{P}_{c_{1}b_{1}}\right) ,  \label{Eq97}
\end{equation}%
and%
\begin{equation}
i\stackrel{.}{S}_{1}S_{1}^{-1}=-\stackrel{.}{\alpha }_{1}\left( \widehat{P}%
_{b_{1}d_{1}}+\widehat{P}_{d_{1}b_{1}}\right) .  \label{Eq98}
\end{equation}%
Here, for simplicity, we set the phases [see Eq. (\ref{Eq1}),(\ref{Eq4})] $%
\varphi _{1}$ and $\varphi _{2}$ equal to zero. The development coefficients
$C_{d_{1}}$, $C_{b_{1}}$, $C_{c_{1}}$ of the state vector $\left| \Phi
_{d_{1}b_{1}c_{1}}\right\rangle $ in the $d_{1}b_{1}c_{1}$--basis satisfy
equations that are similar to equations (\ref{Eq27})--(\ref{Eq29}) if the
substitution $Z_{1}=-C_{d_{1}}$, $Y_{1}=iC_{b_{1}}$ and $X_{1}=C_{c_{1}}$ is
made. These equations are%
\begin{equation}
\stackrel{.}{Z}_{1}=-\stackrel{.}{\alpha }_{1}Y_{1},  \label{Eq99}
\end{equation}%
\begin{equation}
\stackrel{.}{Y}_{1}=-\widetilde{B}X_{1}+\stackrel{.}{\alpha }_{1}Z_{1},
\label{Eq100}
\end{equation}%
\begin{equation}
\stackrel{.}{X}_{1}=\widetilde{B}Y_{1}.  \label{Eq101}
\end{equation}%
The derivative of the mixing parameter $\alpha _{1}$ has an explicit form%
\begin{equation}
\stackrel{.}{\alpha }_{1}=\frac{\lambda }{V_{0}}\frac{1}{1+\left( \frac{%
\lambda t}{V_{0}}\right) ^{2}}.  \label{Eq102}
\end{equation}%
This function has a bell shape with maximum value $\lambda /V_{0}$ at $t=0$.
The modified Rabi frequency $\widetilde{B}=2V_{0}\sqrt{1+\left( \lambda
t/V_{0}\right) ^{2}}$ has a minimum value $2V_{0}$ at $t=0$ and increases to
infinity at $t=\pm \infty $. Qualitatively, the time dependence of the
interaction parameters resembles what we have for the STIRAP induced by the
secant hyperbolic pulses, i.e., $B(t)$ and $\stackrel{.}{\alpha }(t)$.
Therefore, we can apply the same procedure of the calculation of the
nonadiabatic correction and the same approximations.

First, we approximate the Rabi frequency by its expansion in a power series
of $\left( \lambda t/V_{0}\right) ^{2}$, retaining only the first two terms%
\begin{equation}
\widetilde{B}(t)\approx \widetilde{\beta }_{0}\left( 1+\widetilde{g}%
t^{2}\right) ,  \label{Eq103}
\end{equation}%
where $\widetilde{\beta }_{0}=2V_{0}$ and $\widetilde{g}=(\lambda
/V_{0})^{2}/2$ [see Eq. (\ref{Eq66})]. Rescaling time as $\tau =\lambda
t/V_{0}$, we can define the nonadiabatic correction $f_{cA}(+\infty )$ as
follows%
\begin{equation}
f_{cA}(+\infty )=\int_{-\infty }^{+\infty }\frac{\cos \Omega (\tau )}{1+\tau
^{2}}d\tau ,  \label{Eq104}
\end{equation}%
where%
\begin{equation}
\Omega (\tau )=2\eta (\tau +\frac{g_{1}}{3}\tau ^{3}),  \label{Eq105}
\end{equation}%
and $\eta =V_{0}^{2}/\lambda $, $g_{1}=1/2$ [compare the function $\Omega
(\tau )$ with the one in Eq. (\ref{Eq67})]. The Fourier transform%
\begin{equation}
a_{1}(\omega )=\int_{-\infty }^{+\infty }d\tau \stackrel{.}{\alpha }%
_{1}(\tau )e^{-i\omega \tau },  \label{Eq106}
\end{equation}%
of the coupling parameter $\stackrel{.}{\alpha }_{1}(\tau )$ is%
\begin{equation}
a_{1}(\omega )=\pi \exp \left( -\left| \omega \right| \right) .
\label{Eq107}
\end{equation}%
Following the same procedure as the one described in Section VII, we obtain%
\begin{equation}
f_{cA}(+\infty )=\int_{-\infty }^{+\infty }a_{1}(\omega )\eta ^{-\frac{1}{3}%
}Ai\left[ \eta ^{-\frac{1}{3}}\left( 2\eta +\omega \right) \right] d\omega .
\label{Eq108}
\end{equation}%
Employing the same approximation of the Airy function by equation (\ref{Eq72}%
) for positive arguments and taking into account only the negative wing of
the function $a_{1}(\omega )$, we limit the integration in Eq. (\ref{Eq108})
by the boundaries $(-2\eta ,0)$. These boundaries are taken to reduce the
overestimation of the integrand introduced by the approximation applied to
the Airy function and $a_{1}(\omega )$ function. The arguments justifying
this procedure are absolutely similar to those that are applied in Section
IX to the case of the secant hyperbolic pulses.

After some algebra, we obtain%
\begin{equation}
f_{cA}(+\infty )=\pi Q_{1}\exp \left( -\frac{5}{3}\eta \right) ,
\label{Eq109}
\end{equation}%
where%
\begin{equation}
Q_{1}=\frac{1}{2}\left[
%TCIMACRO{\func{erf}}%
%BeginExpansion
\mathop{\rm erf}%
%EndExpansion
(h_{\max })+%
%TCIMACRO{\func{erf}}%
%BeginExpansion
\mathop{\rm erf}%
%EndExpansion
(h_{\min })\right] ,  \label{Eq110}
\end{equation}%
and%
\begin{equation}
h_{\max }=\sqrt{2\eta \left[ \frac{2\sqrt{2}}{3}-\left( \frac{2}{3}\right) ^{%
\frac{4}{3}}+\frac{1}{6}\right] },  \label{Eq111}
\end{equation}%
\begin{equation}
h_{\min }=\sqrt{\frac{\eta }{3}}.  \label{Eq112}
\end{equation}%
This result can be obtained from Eqs. (\ref{Eq86})--(\ref{Eq89}) by the
straightforward substitution $R=1$, $\beta _{0}=2\eta $ and $g=g_{1}$.

We take $Z_{1}(-\infty )=1$ or $C_{d_{1}}(-\infty )=-1$ and $C_{d}(-\infty
)=1$ as initial condition, which means that the reciprocal two-level system
is initially in the ground state $g$. The final state is defined by the
relation $C_{d}(+\infty )=C_{d_{1}}(+\infty )=-Z_{1}(+\infty ).$ Since $%
Z_{1} $ keeps its sign, $C_{d}$ changes sign. The population difference of
the equivalent two-level system $w=\rho _{gg}-\rho _{ee}$ is related to $%
C_{d}$ as $w=C_{d}$. If $C_{d}$ becomes negative, then $\rho _{ee}>\rho
_{gg} $. In this case, to estimate the population fraction left in the
ground state $g$, we have to use the expression%
\begin{equation}
\rho _{gg}(+\infty )=\frac{1+C_{d}(+\infty )}{2}=\frac{1-Z_{1}(+\infty )}{2}.
\label{Eq113}
\end{equation}%
After the substitution $Z_{1}(+\infty )=\sqrt{1-f_{c}^{2}(+\infty )}$ into
Eq. (\ref{Eq113}) and expanding the square root in a power series, we find
the nonadiabatic correction for the Rosen-Zener model%
\begin{equation}
\rho _{gg}(+\infty )\approx \frac{1}{4}f_{cA}^{2}(+\infty )=\frac{\pi
^{2}Q_{1}^{2}}{4}\exp \left( -\frac{10}{3}\eta \right) .  \label{Eq114}
\end{equation}%
Figure 8 shows the comparison of our approximate solution, Eq. (\ref{Eq114}%
), with the exact solution of the Landau-Zener model%
\begin{equation}
\rho _{gg}(+\infty )=\exp \left( -\pi \eta \right) .  \label{Eq115}
\end{equation}%
The difference is small and it originates from the approximations made. To
estimate the accuracy of our result, we compare Eqs. (\ref{Eq114}) and (\ref%
{Eq115}) in logarithmic scale as follows%
\begin{equation}
\ln \left[ \rho _{ee}(+\infty )\right] =-\eta \pi \approx -\eta \left[ \frac{%
10}{3}-\frac{2}{\eta }\ln \left( \frac{\pi Q_{1}}{2}\right) \right] .
\label{Eq116}
\end{equation}%
For large $\eta $, the accuracy of the exponential factor calculation,
expressed in the square brackets in Eq. (\ref{Eq116}), is 6\%.

\newpage

\section{Figure captions}

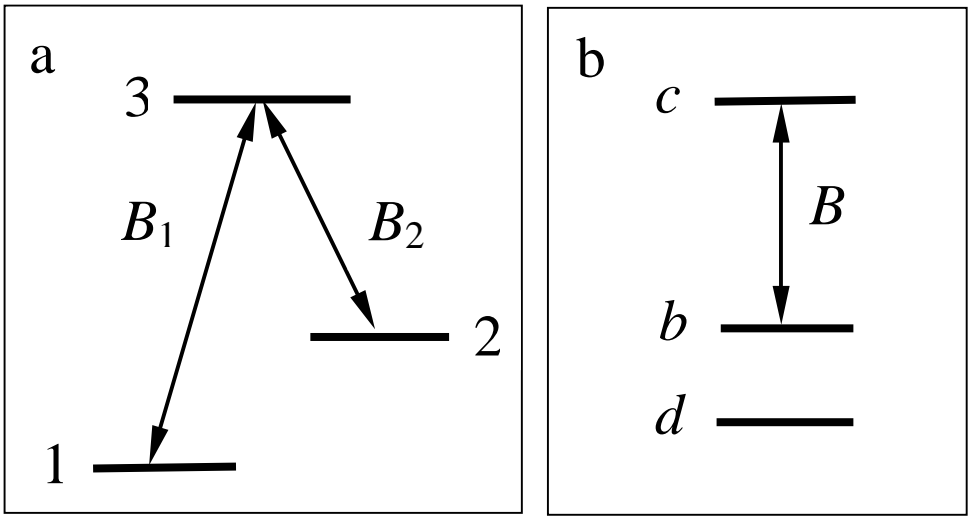
Fig.1. The excitation scheme of the three-level atom by two resonant fields
with interaction constants $B_{1}$ and $B_{2}$: $\Lambda $-scheme (a), and
the same excitation scheme in the basis of dark $d$, bright $b$ and common $%
c $ states (b). $B=\sqrt{B_{1}^{2}+B_{2}^{2}}$ is the generalized
interaction constant for the Raman excitation. The vertical scale in diagram
b is not the energy and represents the initial population of the levels
before the excitation. The higher populated level is put at the bottom and
the lower or unpopulated level is at the top.

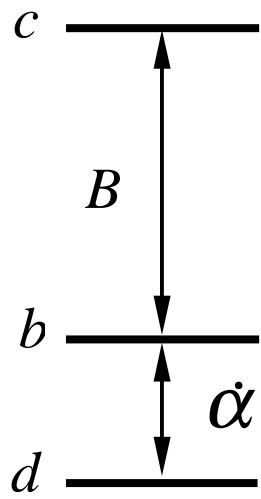
Fig. 2. The bichromatic excitation scheme of the three-level atom in the $%
dbc $ basis if the state developments of $\left| d\right\rangle $ and $%
\left| b\right\rangle $ change in time. The coupling parameter of states $d$
and $b$ is $\stackrel{.}{\alpha }$, the derivative of the mixing angle (see
the text).

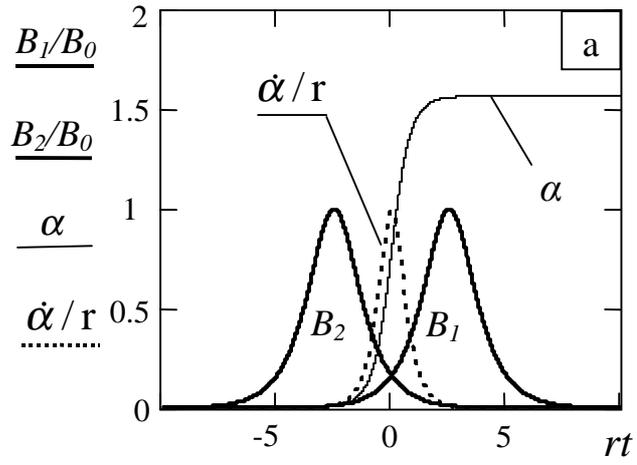
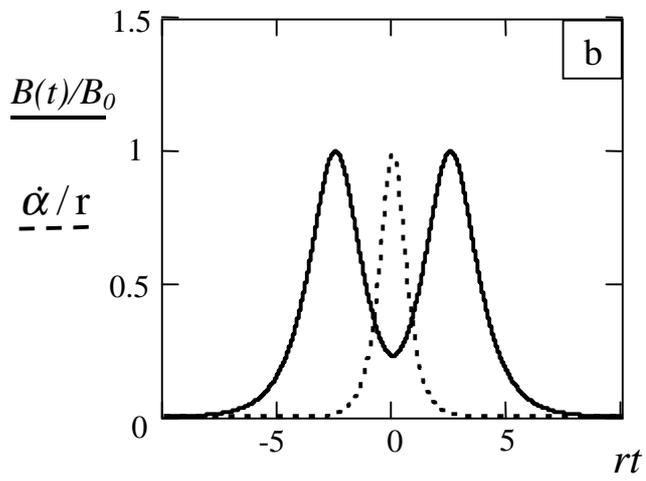
Fig. 3. (a) The pulse train with corresponding interaction parameters $%
B_{1}(t)$ and $B_{2}(t)$ evolving in time (bold lines). They are normalized
by the maximum value $B_{0}$. The delay between pulses is $r\tau =5$, $%
t_{0}=0$. The time is scaled in units of $r$. The time dependence of the
mixing parameter $\alpha $ is shown by the thin line. The dashed line shows
the dependence of the mixing parameter derivative $\stackrel{.}{\alpha }$
normalized by $r$. In (b), the time dependence of the bichromatic Rabi
frequency $\chi /2=B(t)$ (solid line) and the mixing parameter derivative
(dashed line) are shown for comparison.

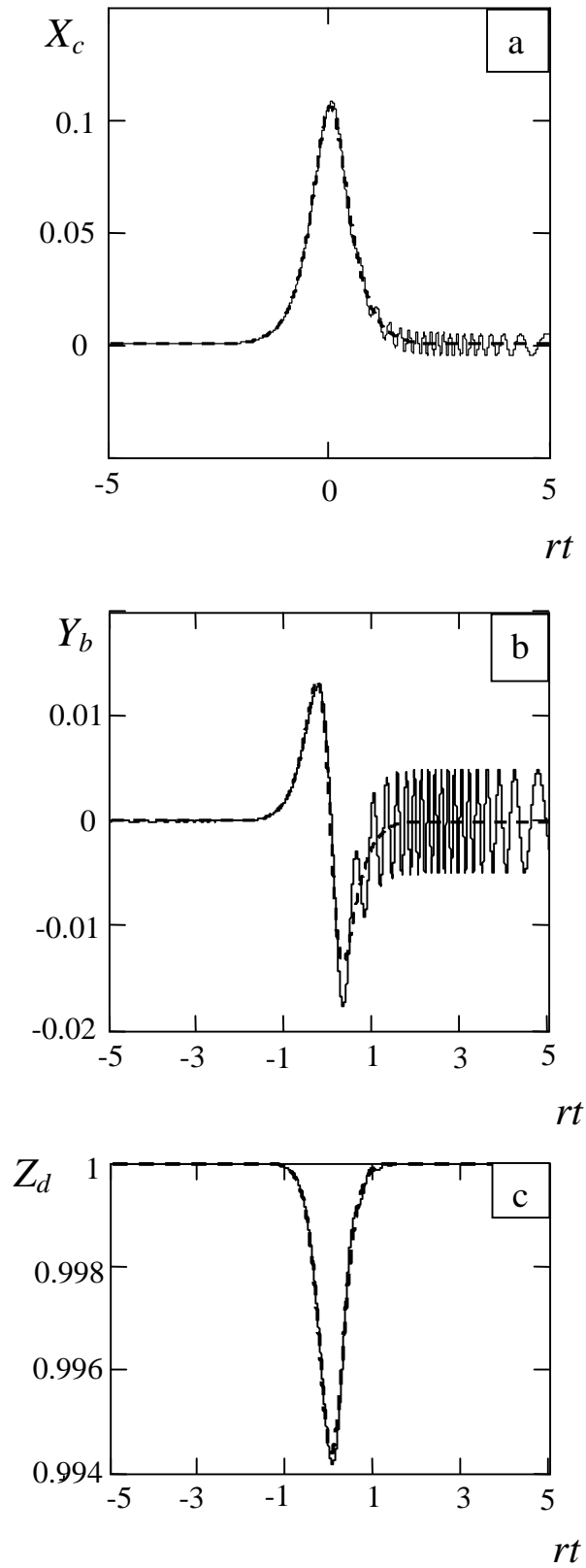
Fig. 4. (a) The evolution of the amplitudes $X_{c}$ (plot a), $Y_{b}$ (plot
b), $Z_{d}$ (plot c) for a pulse train $B_{1}(t)$, $B_{2}(t)$ with the
parameters $B_{0}=42.8r$, $t_{1}=2.5/r$, $t_{2}=-2.5/r$. Solid lines are the
numerical solution of equations (\ref{Eq27})--(\ref{Eq29}). Dashed lines are
analytical approximations given by the first two terms of equations (\ref%
{Eq46},\ref{Eq47}) and the first two terms in each of the parentheses of
equation (\ref{Eq55}).

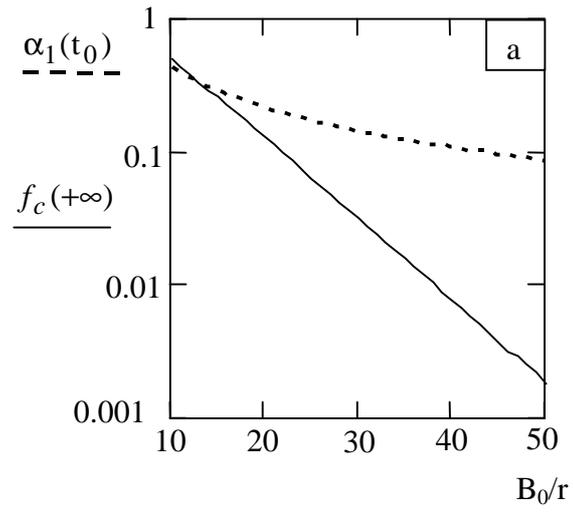
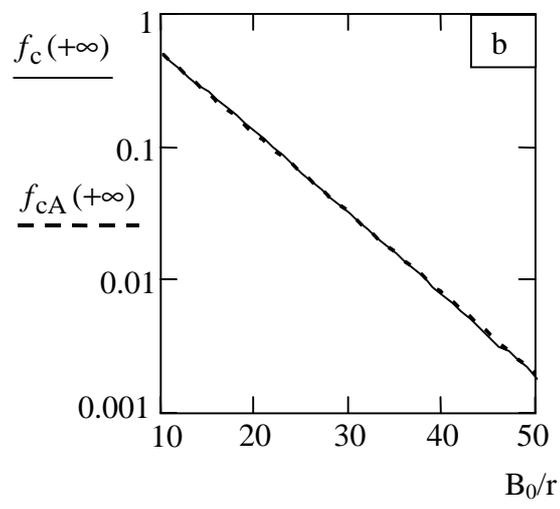
Fig. 5. (a) Comparison of the dependence of the adiabatic $\alpha
_{1}(t_{0}) $ (dashed line) and nonadiabatic $f_{c}(+\infty )$\ (solid line)
parts of the analytical solution of Eqs. (\ref{Eq27})--(\ref{Eq29}) on the
maximum pulse amplitudes $B_{0}$. In (b), the comparison of the nonadiabatic
part, numerically calculated using the actual Rabi frequency $B(t)$ (solid
line) and its parabolic approximation, Eq. (\ref{Eq66}) (dashed line), are
shown. $f_{c}(+\infty )$ is the true nonadiabatic part and $f_{cA}(+\infty )$
is an approximation.

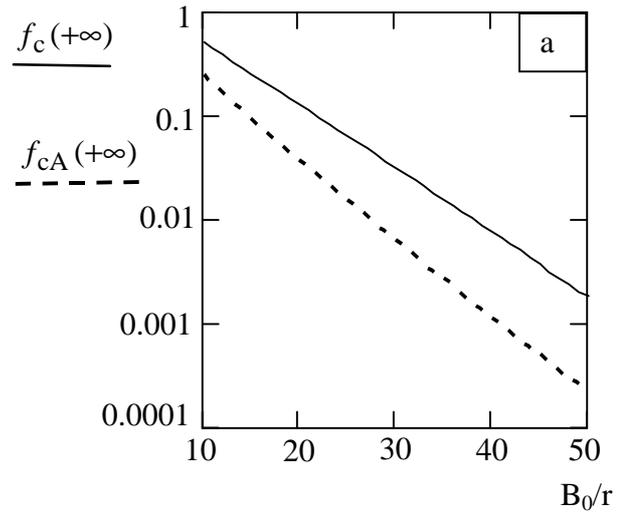
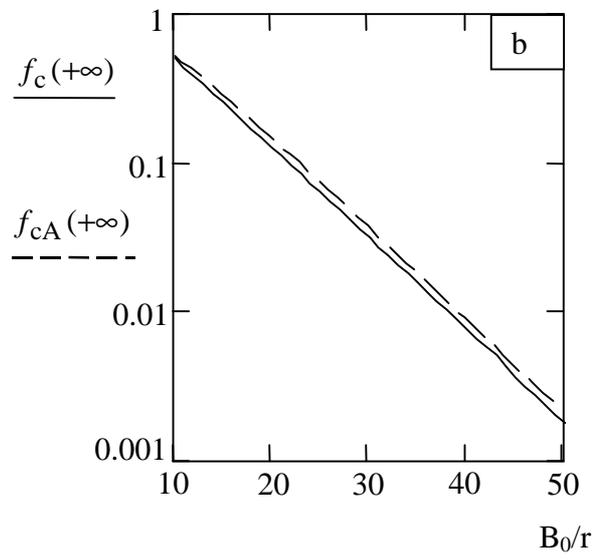
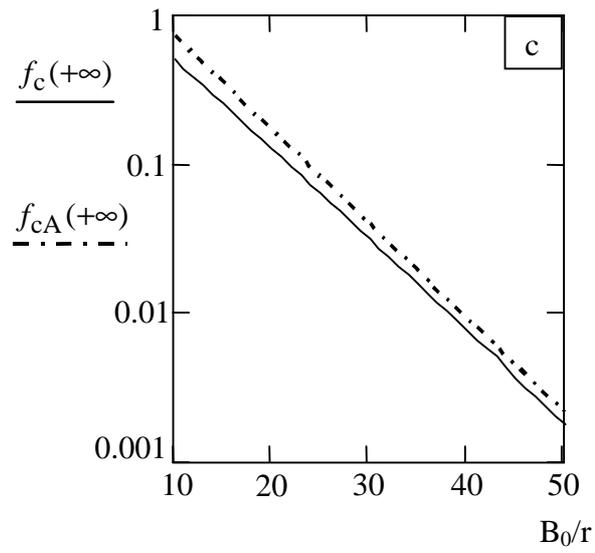
Fig. 6. (a) Comparison of the dependencies of the nonadiabatic contributions
$f_{c}(+\infty )$ (solid line) and $f_{cA}(+\infty )$ (dashed line) versus
the pulse amplitude $B_{0}$. The first is calculated with and the second
without taking into account the time dependence of the mixing parameter
derivative $\stackrel{.}{\alpha }(t)$. (b) The plots of the true, $%
f_{c}(+\infty )$, and the approximate, $f_{cA}(+\infty )$, nonadiabatic
contributions versus $B_{0}$, where $f_{cA}(+\infty )$ (long dashed line) is
calculated taking into account the time dependence of $\stackrel{.}{\alpha }%
(t)$ using the approximation described in the text. (c) The same plots as in
(b), except $f_{cA}(+\infty )$ (dot dashed line), where the correction
factor $Q$ is dropped (see the text).

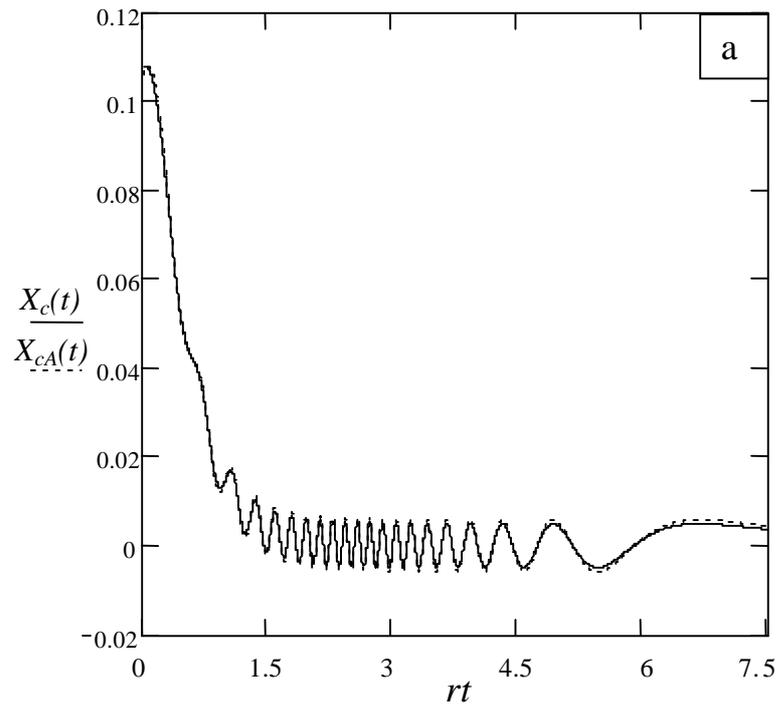
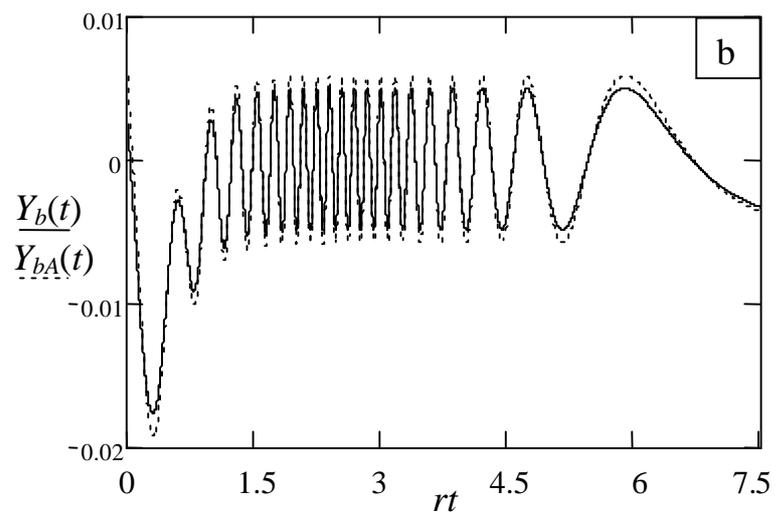
Fig. 7. Numerical solution of the Schr\"{o}dinger equation (solid lines) for
the amplitudes $X_{c}(t)$ (plot a), $Y_{b}(t)$ (plot b) and the
approximation given by equations (\ref{Eq86})--(\ref{Eq87}) (dashed lines)
for the amplitudes $X_{cA}(t)$ (plot a), $Y_{bA}(t)$ (plot b).

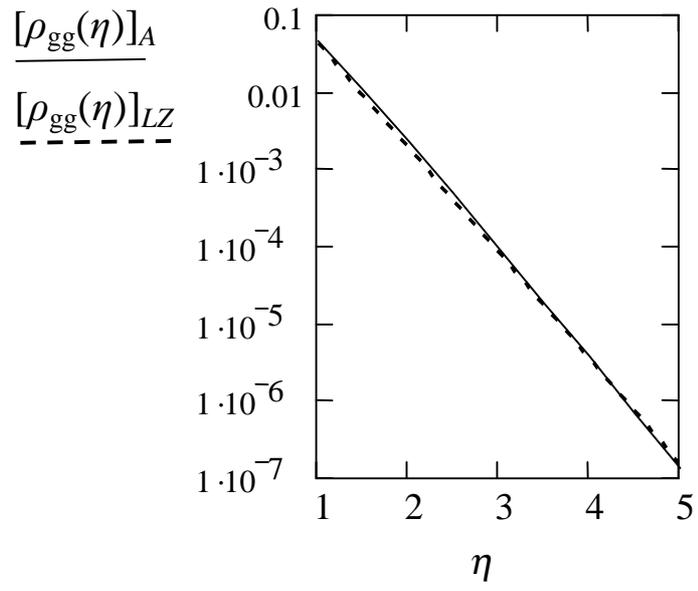
Fig. 8. Comparison of the exact solution of the Landau-Zener model (dashed
line) with our approximate solution (solid line) for the ground state
population of the reciprocal two-level system after level crossing. The
dependence of the exact solution of the Landau-Zener model for the ground
state population, $[\rho _{gg}(\eta )]_{LZ}$, on the parameter $\eta
=V_{0}^{2}/\lambda $ is shown by the dashed line [see Eq. (\ref{Eq115})].
The same dependence of our approximate solution, $[\rho _{gg}(\eta )]_{A}$,
is shown by the solid line. Our solution is given by Eq. (\ref{Eq114}).

\end{document}